\documentclass[showpacs,preprintnumbers,amsmath,amssymb]{revtex4}
\usepackage{calc,capt-of,ifthen}
\usepackage[T1]{fontenc}

\usepackage{graphicx}
\usepackage{dcolumn}
\usepackage{bm}
\usepackage{amsmath}
\usepackage{amsfonts}
\usepackage{amssymb}
\usepackage{ulem}
\usepackage{color}
\usepackage{slashed}
\newcommand{\tmabbr}[1]{#1}
\newcommand{\tmop}[1]{\ensuremath{\operatorname{#1}}}

\definecolor{grey}{rgb}{0.75,0.75,0.75}
\definecolor{orange}{rgb}{1.0,0.5,0.5}
\definecolor{brown}{rgb}{0.5,0.25,0.0}
\definecolor{pink}{rgb}{1.0,0.5,0.5}
\newcommand{\tmfloatcontents}{}
\newlength{\tmfloatwidth}
\newcommand{\tmfloat}[5]{
\renewcommand{\tmfloatcontents}{#4}
  \setlength{\tmfloatwidth}{\widthof{\tmfloatcontents}+1in}
  \ifthenelse{\equal{#2}{small}}
    {\ifthenelse{\lengthtest{\tmfloatwidth > \linewidth}}
      {\setlength{\tmfloatwidth}{\linewidth}}{}}
    {\setlength{\tmfloatwidth}{\linewidth}}
  \begin{minipage}[#1]{\tmfloatwidth}
    \begin{center}
      \tmfloatcontents
      \captionof{#3}{#5}
    \end{center}
  \end{minipage}}

\newcommand{\beq}{\begin{equation}}
\newcommand{\eeq}{\end{equation}}
\newcommand{\bea}{\begin{eqnarray}}
\newcommand{\eea}{\end{eqnarray}}

\newcommand{\Angs}{\text{\hspace{.2em}\r{A}}}
\bmdefine\kvec{k}
\newcommand{\aone}{\bm{a}_1}
\newcommand{\atwo}{\bm{a}_2}
\newcommand{\I}{\mathrm{i}}

\begin{document}

\title{Charge Transport in Disordered Graphene-Based Low Dimensional Materials}

\author{Alessandro Cresti$^{1,2}$, Norbert Nemec$^3$, Blanca Biel$^{1,2}$, Gabriel Niebler$^{4,5}$, Fran{\c{c}}ois Triozon$^1$, Gianaurelio Cuniberti$^4$ and Stephan Roche$^2$}

\affiliation{
$^1$ CEA, LETI-Minatec, 17 rue des Martyrs, 38054 Grenoble Cedex 9, France\\
$^2$ CEA, Institute for Nanoscience and Cryogenics, INAC/SPSMS/GT, 17 rue des Martyrs, 38054 Grenoble Cedex 9, France\\
$^3$ Theory of Condensed Matter Group, Cavendish Laboratory, University of Cambridge, England\\
$^4$ Institute for Materials Science, TU Dresden, D-01062 Dresden, Germany \\
$^5$ Department of Condensed Matter Physics, Faculty of Mathematics and  Physics, Charles University, Ke Karlovu 5, 121 16 Prague 2, Czechia
}

\begin{abstract}
Two-dimensional graphene, carbon nanotubes and graphene nanoribbons represent a novel class of low dimensional materials that could serve as building blocks for future carbon-based nanoelectronics. Although these systems share a similar underlying electronic structure, whose exact details depend on confinement effects, crucial differences emerge when disorder comes into play. In this short review, we consider the transport properties of these materials, with particular emphasis to the case of graphene nanoribbons. After summarizing the electronic and transport properties of defect-free systems, we focus on the effects of a model disorder potential (Anderson-type), and illustrate how transport properties are sensitive to the underlying symmetry. We provide analytical expressions for the elastic mean free path of carbon nanotubes and graphene nanoribbons, and discuss the onset of weak and strong localization regimes, which are genuinely dependent on the transport dimensionality. We also consider the effects of edge disorder and roughness for graphene nanoribbons in relation to their armchair or zigzag orientation. 
\end{abstract}

\maketitle

\tableofcontents

\section{Introduction}
Since the discovery of fullerenes ($C_{60}$), carbon-based materials have been the subject of intense research, which led to the discovery of carbon nanotubes \cite{1,Loiseau:book06,nanotubes,RMP} and the fabrication of individual one-atom thick graphene layers \cite{graphene}. This opens unprecedented avenues for the investigation of quantum transport in low dimensional 1D and 2D systems, as well as attracting the interest of industries, given the potential for innovative applications. Carbon nanotubes science is now a mature field of research, whose theoretical foundations have been reviewed in many textbooks (see \cite{nanotubes,RMP} and references therein). More recently, the possibility to single out a graphene plane, either through an exfoliation process \cite{graphene}, or by means of an epitaxial growth mechanism \cite{BER_SCI312}, has allowed the whole community of mesoscopic physics to revisit basic foundations of quantum transport in 2D systems, such as quantum Hall effects or weak localization phenomena \cite{ZHA_N438,ZHA_PRL96,GEI_NM6,OEZ_APL91,OEZ_PRL99,JIA_PRL99,NOV_SCI315,WU_PRL98}, in a material with remarkable electronic properties \cite{RMP}. Additionally, the routes for an alternative carbon-based nanoelectronics are actively investigated ranging from device optimization to graphene integration at the wafer scale \cite{DAY,LEM_IEEE28,HAN_PRL98,CHE_PE40,ECH_EPJ148,MOR_PRL100,BER_SCI312}. However, since truly two-dimensional graphene is a zero-gap semiconductor, its use as an active electronic device, such as a field effect transistor, requires the reduction of its lateral size to benefit from induced quantum confinement effects. Energy band-gap engineering has been demonstrated with both types of materials, carbon nanotubes (CNTs) and graphene nanoribbons (GNRs) \cite{JAV_N424,JAV_NL4,HAN_PRL98}. In the case of GNRs-based field effect transistor, this results in an increase of performances with downscaling the ribbon width from several tens of nanometers to 2 nm \cite{LEM_IEEE28,HAN_PRL98,CHE_PE40,ECH_EPJ148,MOR_PRL100,XIA_SCI319}. In contrast to carbon nanotubes, whose integration in operating devices at a large scale remains a tremendously complicated challenge, GNRs present the potential to be fabricated and massively integrated in complex architectures, thanks to the use of conventional lithographic techniques.

In this work, we discuss the transport properties of carbon-based low dimensional materials by focusing on the effect of a case of disorder potential, namely the Anderson type model \cite{Anderson1958}. This model has been widely employed in studying the scaling theory of localization, and through analytical and numerical results, it allows us to overview the behavior of the basic transport length scales (elastic mean free path and localization lengths) and associated transport regimes in systems with different symmetries and transport dimensionality. In the case of graphene nanoribbons, the presence of the edges exposes the system to further sources of disorder. In order to account for this peculiarity, we also consider the effects of impurities and roughness at the edges of the GNRs on the transport properties. 

Section \ref{section_cleansystem} is devoted to an overview of the electronic structure and quantum transport properties of 2D-graphene, quasi-one-dimensional carbon nanotubes and graphene nanoribbons in the absence of disorder. We start from a simple nearest-neighbors (n-n) tight-binding model, and highlight differences and similarities between these different carbon structures due to their different dimensions and lack or presence of edges, as these are the underlying factors that determine the specific response of their transport properties to disorder. We include a discussion about how the electronic structure of narrow graphene nanoribbons is modified when performing {\it ab initio} calculations and how to mimic this behavior by means of a third nearest-neighbors tight-binding model. In the case of nanoribbons, we also elaborate on the dependence of spatial symmetry of the system on the armchair or zigzag edge orientation and the number of carbon atoms in the primitive cell, since in certain cases, this has a deep impact on the transport features. At the end of the section, two interesting examples of transport in clean nanoribbons are also analyzed. The first one is the so called "pseudodiffusive" transport regime, which presents curious and intriguing analogies with the diffusive transport regime of disordered metals. The second one explores the effects of high magnetic fields on the energy bands, the transport properties and on the spatial distribution of electrical currents flowing through the ribbons.   

In section \ref{section_disorder}, we consider the effect of disorder on the density of states and on the transport properties of the carbon nanotubes and nanoribbons. We focus on the analysis of typical length scales such as the mean free path and the localization length, taking an Anderson-type disorder as a reference model. In spite of its simplicity, this type of disorder allows us to point out the differences between 2D graphene and other graphene-derived materials. More realistic types of disorder would lead to more complicated effects that might mask the essential dependence of the transport properties on the dimensionality and symmetries of the investigated systems. A subsection is also devoted to the edge disorder (roughness and Anderson-type) in graphene nanoribbons. The effect of this kind of disorder on the transport properties very critically depends on the armchair or zigzag edge orientation of the ribbons. In the literature, several different models of edge disorder are considered and investigated with not always unanimous conclusions. In the last part of the section, we discuss the charge mobility in graphene based systems subjected to different types of disorder, since such quantity is key to assess the performances of device performances. 

Section \ref{conclusions} concludes the report by stressing the most relevant features.

\section{\label{section_cleansystem}Electronic properties and transport in defect-free systems}
In most cases, the energetics of graphene based materials can be conveniently described as a first approach by a $\pi$-effective electron model \cite{RMP}. Indeed, for the honeycomb geometry, $\sigma$ bonds are formed by three out of four valence electrons ($sp^{2}$-hybridization), whereas the remaining single electron per carbon atom occupies a $2p_{z}$ orbital whose hybridization with first neighbors generate $\pi$ and $\pi^{*}$ bands. A restriction of the Hamiltonian to a single $p_{z}$ orbital per carbon atom provides thus an effective energetic model that well describes the electronic properties close to the charge neutrality point (CNP). In the case of graphene nanoribbons, we assume that the $\sigma$ bonds at edges are saturated with H atoms, thus eliminating dangling bonds. 

\subsection{\label{section_2D}2D graphene band structure}
CNP locates the Fermi level for 2D undoped graphene and metallic nanotubes and graphene ribbons, and the mid-gap of semiconducting CNTs and GNRs. Defining $\gamma_{0}$ as the integral overlap between $p_{z}$ orbitals, the electronic properties of 2D graphene are straightforwardly derived by diagonalizing a $2\times2$ matrix as

\begin{figure}[ht]
 \centering \includegraphics[width=8cm]{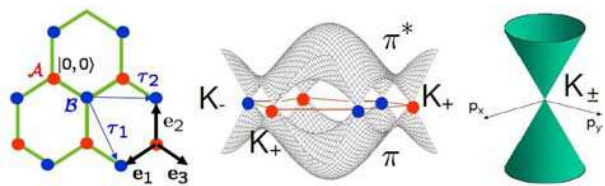}
 \caption{Honeycomb lattice, graphene $\pi$ and $\pi^{*}$ bands, and dispersion relation close to the ${\bf K}_{\pm}$ points of the first Brillouin zone.}
 \label{FIG2}
\end{figure}

\begin{equation}
    \mathbf{H}(\mathbf{k})=\left(\begin{array}{cc}
  0 & f(\mathbf{k})\\
  f^{*}(\mathbf{k}) & 0
\end{array}\right)
\end{equation}
with
$f(\mathbf{k})=\gamma_{0}\sum_{m}e^{i\mathbf{k}.\mathbf{e}_{m}}$ ($\mathbf{e}_{m}$ are the vectors that connect A-types site to the three nearest neighbor B-type sites, see Figure \ref{FIG2}, left panel) and $\gamma_0\approx -2.7$ eV. It is readily shown that
\beq \label{eq:2D_bands_1st_NN}
E_{\pm}(k_x,k_y)=\pm\gamma_{0}\bigl[3+4\cos(\frac{\sqrt{3}k_{x}a}{2})\cos(\frac{k_{y}a}{2})+2\cos(k_{y}a)\bigr]^{1/2} \eeq
(see Figure \ref{FIG2}, central panel).

 The underlying symmetry and sublattice degeneracy of the honeycomb lattice can be unveiled from a different starting point. In a tight-binding representation, one can separate the two sublattices as
\beq
\hat{\cal H}=\gamma_{0}\sum_{\displaystyle {\bf R}\in B.L.}\sum_{\displaystyle m=1}^{3}\hat{\cal B}^{\dagger}_{\bf R+e_{m}}
\hat{\cal A}_{\bf R}+h.c,
\eeq
by introducing creation ($\hat{\cal A}^{\dagger},\hat{\cal B}^{\dagger}$) and annihilation ($\hat{\cal A},\hat{\cal B}$) operators for electrons on site A or B, respectively. To diagonalize the Hamiltonian, one applies Bloch theorem to the Bravais lattice, which yields
\beq
  \hat{\cal A}^{\dagger}_{\bf R} =\displaystyle \frac{1}{\sqrt{N_{m}}}{\displaystyle \sum_{k\in 1BZ}} e^{-i{\bf k}.{\bf R}}\hat{\cal A}^{\dagger}_{\bf k} \ \ \ \ \ \ \ \ \ \ \
  \hat{\cal B}^{\dagger}_{\bf R+e_{m}}=\displaystyle\frac{1}{\sqrt{N_{m}}}{\displaystyle \sum_{k\in 1BZ}}e^{-i{\bf k}.({\bf R}+\mathbf{e}_{m})}\hat{\cal B}^{\dagger}_{\bf k}
\eeq
with $N_{m}$ the number of unit cells. By means of this transformation
\begin{eqnarray}
  \hat{\cal H}={\displaystyle \sum_{k\in 1BZ}}h_{k}\hat{\cal B}^{\dagger}_{\bf k}\hat{\cal A}_{\bf k},
\end{eqnarray}
with
\begin{eqnarray}
  h_{k}=\gamma_{0}{\displaystyle \sum_{m=1}^{3}e^{-i{\bf k}.{\bf e}_{m}}}
  =|h_{k}|e^{i\theta_{k}}.
\end{eqnarray}
In the sublattice basis, at a given crystal momentum, this also reads
\beq\hat{\cal H}={\displaystyle \sum_{k\in 1BZ}} (\hat{\cal A}^{\dagger}_{k},\hat{\cal B}^{\dagger}_{k})
  \left(\begin{array}{cc}
       0 & h_{k}\\
       h_{k} & 0
  \end{array}\right)
  \left(\begin{array}{c}
     \hat{\cal A}_{k} \\
     \hat{\cal B}_{k}
\end{array}\right) \ .
\eeq
Now, by performing a linearization close to $\mathbf{K}$ points, i.e. $\mathbf{k}=\mathbf{K}_{\pm}+\mathbf{p}/\hbar$ at first order in ${\mathbf{p}}=-i\hbar{\mathbf{\nabla}}$, one finds
\beq{\cal H}_{K_{+}}=v_{F}\left(\begin{array}{cc}
   0&p_{x}+ip_{y}\\
   p_{x}-ip_{y}&0\\
\end{array}\right) \ ,\eeq
where $v_{F}=\sqrt{3}a|\gamma_{0}|/2\hbar$ is the Fermi velocity.
Therefore, the effective Hamiltonian takes the form of a Dirac Hamiltonian for massless particles
\beq
{\cal H}_{K_{+}}(\mathbf{p}) = v_{F}{\mathbf{\sigma}}.\mathbf{p} \ \ \ \ \ \ \ \ \ \ \
  {\cal H}_{K_{-}}(\mathbf{p}) = -v_{F}{\mathbf{\sigma}}^{*}.\mathbf{p} \ ,
\eeq
with $\sigma_{x,y,z}$ the Pauli matrices. Finally, by identifying each of the $\mathbf{K}$-points with the index, these Hamiltonians can be rotated into a diagonal form with the unitary operator ($\theta=\arctan(p_{y}/p_{x})$) 
\beq{\cal U}_{\alpha}(\mathbf{p})=\frac{1}{\sqrt{2}}
     \left(\begin{array}{cc}
         1&1\\
         -\alpha e^{-i\alpha\theta}&\alpha e^{-i\alpha\theta}\\
\end{array}\right) \ , \eeq
so that
\beq {\cal U}^{\dagger}_{\alpha}(\mathbf{p}){\cal H}_{\alpha}(\mathbf{p}) {\cal U}^{\dagger}_{\alpha}(\mathbf{p})
      =\alpha v_{F}\left(\begin{array}{cc}
        -|\mathbf{p}|&0\\
        0& |\mathbf{p}| \\
\end{array}\right)=-\alpha v_{F}|\mathbf{p}|\sigma_{z} \ . \eeq
The dispersion relation describes a cone-type structure in the reciprocal space (see Figure \ref{FIG2}, right panel), whereas the corresponding eigenvectors define a spinor (pseudospin) \cite{RMP}
\beq|\Psi_{\mathbf{p}}^{+}\rangle=\frac{1}{\sqrt{2}}
      \left(\begin{array}{c}
      \psi_{p}^{+}(A)\\
      \psi_{p}^{+}(B)
   \end{array}\right)=\frac{1}{\sqrt{2}}
   \left(\begin{array}{c}
      se^{i\theta/2}\\
      e^{-i\theta/2}
\end{array}\right) \ .\eeq
The spinor is also an eigenstate of the helicity operator
\mbox{$\hat{\epsilon}=(1/2)\hat{\mathbf{\sigma}}.{\mathbf{p}}/|\mathbf{p} |$}, such that \mbox{$\hat{\epsilon} |\Psi^{+}_{\mathbf{p}}(s=\pm1)\rangle=\pm (1/2)|\Psi^{+}_{\mathbf{p}}(s)\rangle$}. Thus, eigenstates have a well-defined helicity (good quantum number). Note, however, that the above derivation is valid in first order in $\mathbf{p}$, whereas trigonal warping, which appears at second order \cite{nanotubes}, will bring substantial modifications of band structure with consequences on transport properties.

The electronic properties of the 2D graphene can thus be described by an effective massless Dirac fermion model in the vicinity of the CNP, with linear dispersion and electron-hole symmetry. These properties derived close to $\mathbf{K}$ point are also valid for 1D systems such as metallic nanotubes and wide armchair nanoribbons. However, other symmetries result in semiconducting systems with varying gaps.
Semiconducting nanotubes and ribbons with increasing diameter (or width) show a linear downscaling of their associated energy gaps.
By using proper boundary conditions, the electronic band structure of both types of systems can be analytically derived.

In the presence of a uniform magnetic field $B$ threading the 2D graphene sheet along the orthogonal $z$ direction, we make use of the minimal substitution in the Hamiltonian. If we choose the first Landau gauge, the vector potential is $\mathbf{A}=(By,0,0)$ and the new momentum operator is
\beq \mathbf{\pi}=\left(\begin{array}{c}\pi_x\\[3mm]\pi_y \end{array}\right)=\left(\begin{array}{c}p_x-\displaystyle\frac{e}{c}By\\[3mm]p_y \end{array}\right)\eeq
and so
\beq [\pi_x,\pi_y]=i\frac{\hbar^2}{\ell_0^2} \ \ \ \ \ \ {\rm with}\ \ \ \ \ \ \ell_0=\sqrt{\frac{c\hbar}{eB}} \ .\eeq
Let us define the following operators
\beq \eta=\frac{1}{\sqrt{2}}\frac{\ell_0}{\hbar}(\pi_x+i\pi_y) \ \ \ ; \ \ \ \ \eta^\dag=\frac{1}{\sqrt{2}}\frac{\ell_0}{\hbar}(\pi_x-i\pi_y) \ .\eeq
They can be identified as annihilation and creation operators, since $[\eta,\eta^\dag]=1$. The Hamiltonian around $K_+$ is now
\beq\mathcal{H}_{K_+}=\beta \ \left( \begin{array}{cc} 0 & \eta \\[3mm] \eta^\dag & 0  \end{array} \right) \ \ \ \ \ \ {\rm with} \ \ \ \ \ \ \beta=\sqrt{2} \ \frac{\hbar v_F}{\ell_0} \ . \eeq
Let us consider the square of the Hamiltonian
\beq\mathcal{H}_{K_+}^2=\beta^2 \ \left( \begin{array}{cc} \eta^\dag \eta & 0 \\[3mm] 0 & \eta\eta^\dag  \end{array} \right)=\beta^2 \ \left( \begin{array}{cc} \eta^\dag \eta & 0 \\[3mm] 0 & \eta^\dag\eta+1  \end{array} \right). \eeq
We can identify $\eta^\dag\eta$ as the number operator, therefore the eigenvalues of $\mathcal{H}$ (given by the square root of the eigenvalues of $\mathcal{H}^2$) are \cite{MCC_PR104}
\beq E_n=\pm\beta\sqrt{n}\approx 31.65  \ \sqrt{B}\sqrt{n} \ \ \ \ \ \ {\rm with} \ \ \ \ \ n=0,1,2,3... \ ,\eeq
where the eigenvalues $E_n$ are expressed in meV and the magnetic field B in T.
In contrast to ordinary two-dimensional electron gases, in graphene Landau levels are proportional to the square root on the magnetic field and to the square root of the integer number $n$. Moreover, a double degenerate zero energy level is present.   

\subsection{\label{section_nnTB}Graphene nanoribbons band structure: nearest-neighbor tight-binding model}
Graphene nanonoribbons are strips of graphene that can be obtained by cutting a graphene sheet along a certain direction. 
Depending on this direction, the edges of the ribbon can be armchair-like (Fig.\ref{fig_ribbon_armchair}(a)) or zigzag-like (Fig. \ref{fig_ribbon_zigzag}(a)).
In the following, we will refer to an armchair ribbon composed of $N$ dimers lines as $N$-aGNR, and to a zigzag ribbon composed of $N$ zigzag lines as $N$-zGNR.  

The band structures of ideal GNRs with width below $\sim 100$ nm and well defined edge symmetries (zigzag or armchair types) are dominated by confinement effects and van Hove singularities \cite{NAK_PRB54,WAK_PRB59,WAK_PRB64,EZA_PRB73,PER_PRB73,BreyFertigPRB,MUN_PRB74,WAN_PRB75,ZHE_PRB75}, similarly to carbon nanotubes. As for 2D graphene and carbon nanotubes, a nearest neighbor tight-binding model is found to describe low energy properties with a degree of approximation high enough for many applications. This approach has been widely employed for studying transport properties in pure or defected GNRs \cite{NAK_PRB54,WAK_PRB59,WAK_PRB64,PER_PRB73,MUN_PRB74,ZHE_PRB75,ARE_NL7,GUN_APL90,GUN_PRB75}.

\subsubsection{Armchair nanoribbons}
\begin{figure}[hb]
 \centering \includegraphics{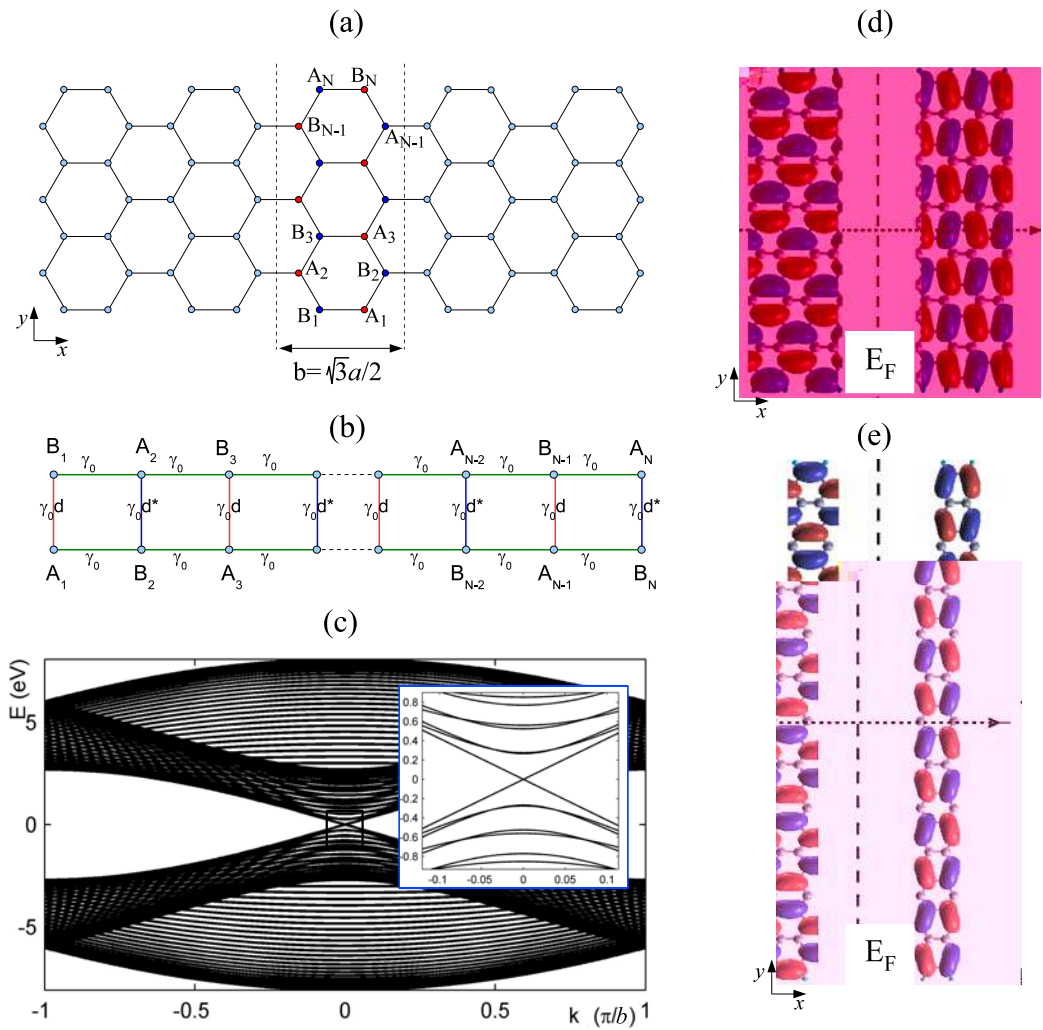}
 \caption{(a) Structure and primitive cell of an armchair ribbon composed of $N$ dimer lines. (b) Equivalent coupled chains with $N$ orbitals each. (c) Energy bands for a metallic ribbon with $N=53$. (d) 2D projection (at the $\Gamma$ point) of the two wavefunctions just below and above the Fermi level for a 20-aGNR. (e) The same for a 35-aGNR. Blue or red colors correspond to opposite sign of the wavefunction.}
 \label{fig_ribbon_armchair}
\end{figure}

An armchair GNR unit cell contains $N$ A-type atoms and $N$ B-type atoms, see Fig. \ref{fig_ribbon_armchair}(a). Thanks to the bipartite lattice of graphene, the total wavefunction of the system can be constructed by a linear combination of A-type $\Psi_{A}$ and B-type $\Psi_{B}$ sublattice wavefunctions. By  applying Dirichlet boundary conditions for the wavefunctions at both edges of the ribbon ($q_{y}=2p\pi/(\sqrt{3}a(N+1))$, $p=1,2,..,N$), one gets \cite{ZHE_PRB75}
\beq
(1/{\cal N})|\Psi\rangle= c_{A}\sum_{j=1}^{N}\sum_{xA_{j}}e^{ik xA_{j}}\sin(\sqrt{3}q_{y}aj/2)|A_{j}\rangle +c_{B}\sum_{j=1}^{N}\sum_{xB_{j}}e^{ik xB_{j}}\sin(\sqrt{3}q_{y}aj/2)|B_{j}\rangle,
\eeq
with ${\cal N}=\sqrt{2/N_{x}(N+1)}$ a normalization factor. Then, by rewriting the Schr\"odinger equation as a $2\times2$ matrix, eigenvalues and eigenvectors are readily found:
\beq
E(k,q_{y})=\pm |\gamma_{0}(2e^{ika/2}\cos(\sqrt{3}q_{y}a/2)+e^{-ika})|=\pm |\Gamma(k,q_{y})|
\eeq
and
\beq
|\Psi\rangle=1/\sqrt{2}\biggl(\Psi_{A}\pm \sqrt{\Gamma^{*}(k,q_{y})/\Gamma(k,q_{y})}\Psi_{B}\biggr).
\eeq

By analyzing $\Gamma(k,q_y)$, we can see that the ribbon is metallic if $N=3m+2$, where $m$ is an integer number, and semiconducting in the other cases. In particular, we can obtain the value evaluate the energy gap $\Delta_N$ as a function of the $N$ dimer chains:
\begin{equation} \label{eqDELTA}\left\{\begin{array}{lcl}
 \Delta_{3m}&=&\displaystyle |\gamma_0| \left(4\cos\frac{\pi m}{3m+1}-2\right)\\[5mm]
 \Delta_{3m+1}&=&\displaystyle |\gamma_0| \left(2-4\cos\frac{\pi (m+1)}{3m+2}\right)\\[3mm]
 \Delta_{3m+2}&=&0
\end{array}\right.\end{equation}
with $\Delta_{3m}>\Delta_{3m+1}>\Delta_{3m+2}=0$. However, as reported in section \ref{abinitio}, {\it ab initio} calculations do not predict the existence of metallic ribbons and a give a different gap size hierarchy, as discussed in Sect.\ref{abinitio}.

Further insight into the electronic structure of graphene nanoribbons can be gained by representing the Hamiltonian on the Bloch sums. If we order the basis as $A_1,B_2,A_3,...,A_{N-1},B_N$ and $B_1,A_2,B_3,...,B_{N-1},A_N$, the Hamiltonian can be split into four $N\times N$ blocks
\begin{equation} \label{eq:HaK}
\mathbf{H}(k)=\gamma_0\left(\begin{array}{cccccccc} 
	0&1&0&...&d&0&0&... \\[3mm]
	1&0&1&...&0&d^*&0&... \\[3mm]
  0&1&0&...&0&0&d&... \\[1mm]
  ...&...&...&...&...&...&...&...\\[1mm]
  d^*&0&0&...&0&1&0&... \\[3mm]
  0&d&0&...&	1&0&1&...\\[3mm]
  0&0&d^*&...&0&1&0&...\\[1mm]
  ...&...&...&...&...&...&...&...
\end{array}\right)
\end{equation}
where $d(k)=e^{ikb/2}$.
The diagonal blocks are tridiagonal matrices with 0 diagonal elements and $\gamma_0$ off-diagonal elements, the off-diagonal blocks are diagonal matrices with alternating $\gamma_0 \exp(\pm i k b/2)$ elements. Hamiltonian (\ref{eq:HaK}) is equivalent to the Hamiltonian of two coupled chains with $N$ orbitals, as indicated in Fig. \ref{fig_ribbon_armchair}(b). Within each chain, the nearest neighbor orbitals are coupled by the hopping parameter $\gamma_0$ and have vanishing onsite energies. The interchain hopping is given by the parameters $\gamma_0 \exp(\pm i k b/2)$. By diagonalizing (\ref{eq:HaK}), the energy bandstucture is obtained. As an example, the energy bands of a metallic armchair ribbon with $N=53$ are reported in Fig. \ref{fig_ribbon_armchair}b. Note the typical Dirac-like linear dispersion around $k=0$. The mapping of the six Dirac points of 2D graphene into $k=0$ can be easily understood by projecting the graphene band structure onto the axis corresponding to the armchair orientation \cite{FUJ_JPSJ65}. 
At $k=0$, the Hamiltonian (\ref{eq:HaK}) is real and our coupled chains turn into a ladder with two legs and $N$ rungs and all hopping parameters equal to $\gamma_0$. The eigenvalues of such a system are analytically known \cite{FAB_PRB46,GOP_PRB49} and give again the energy gap $\Delta_N$ reported in Eq. (\ref{eqDELTA}). 

The n-n tight-binding bandstructure of armchair nanoribbons with $N=3m$, $3m+1$ and $3m+2$ is summarized in Fig. \ref{mosaic} for $N=$9, 10 and 11.

From the viewpoint of spatial symmetry, aGNRs, and GNRs in general, are very different from nanotubes. In fact, an ideal (perfect) carbon nanotube presents a well defined symmetry with respect to a large number of mirror planes containing the tube axis. As a consequence, the parity (even or odd) of the wavefunction has been demonstrated to have an impact on the transport properties of defected carbon nanotubes-based systems, as scattering can only occur between eigenstates with the same parity \cite{CHO_PRL84,KIM_PRB71,KIM_APL88,ZHO_PRB75}. On the contrary, wavefunctions in GNRs do not always present a well defined parity associated to mirror reflections with respect to the axis of the ribbon. The spatial symmetry depends on both the edge termination (zigzag or armchair) and the even or odd number of chains that compose the ribbons. In the case for odd-indexed armchair ribbons (as the 9-aGNR and 11-aGNR in Fig. \ref{mosaic}), the system is invariant under mirror reflection with respect to a plane perpendicular to the ribbon and containing its axis. In the case of even-indexed armchair ribbons (as the 10-aGNR in Fig. \ref{mosaic}), the system is "asymmetric," in the sense that it is no more invariant under the mirror reflection operation. Still, it is invariant under glide plane symmetry, which consists of the mirror reflection followed by a fractional translation along the ribbon axis. Figures \ref{fig_ribbon_armchair}(d,e) show the 2D-projection at the $\Gamma$ point of the two wavefunctions just below and above the Fermi level for perfect asymmetric armchair ribbons (calculated with the SIESTA package \cite{SAN_IJQC65}). As clearly visible in Figs. \ref{fig_ribbon_armchair}(d,e), the existence of a mirror symmetry plane in ribbons leads to a well defined parity of the wavefunctions with respect to that mirror plane. The well defined (or not) parity of the wavefunctions does not have any effect on the electronic transport properties of ideal unperturbed ribbons, but it may have a huge impact in the presence of defects. For example, the electronic transport properties of boron doped armchair ribbons have been proved, by means of \textsl{ab initio} calculations, to strongly depend on the symmetry of the ribbon, as B-induced potentials that preserve the parity of the wavefunctions do not affect the conductance of odd-indexed ribbons at low energies [Biel, B.; Blase, X.; Triozon, F.; Roche, S. {\it submitted}]. This suggests that scattering by certain defects might be suppressed, provided that the defects preserve the underlying symmetric geometry of the ribbon.

\subsubsection{Zigzag nanoribbons}
\begin{figure}[ht]
 \centering \includegraphics{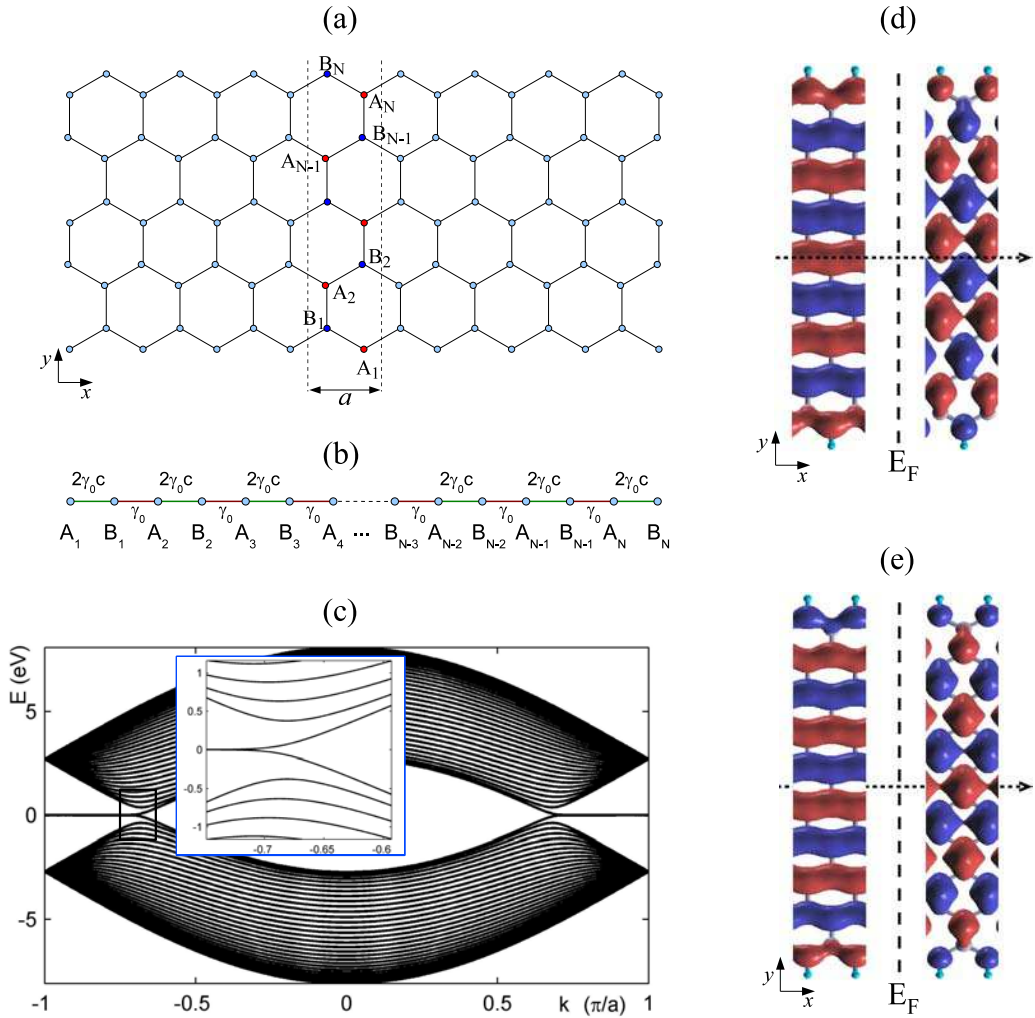}
 \caption{(a) Structure and primitive cell of a zigzag ribbon composed of $N$ carbon lines. (b) Equivalent finite chain with $2N$ orbitals. (c) Energy bands for $N=32$. (d) 2D projection (at the $\Gamma$ point) of the two wavefunctions just below and above the Fermi level for a 9-zGNR. (e) The same for the 10-zGNR. Blue or red colors correspond to opposite sign of the wavefunction.}
 \label{fig_ribbon_zigzag}
\end{figure}

For zigzag GNRs, the handling of the boundary conditions is slightly more involved. A possible approach is solving the Dirac equation requiring that the wavefunction vanishes at the edges of the ribbon \cite{BreyFertigPRB}. As evident from Fig. \ref{fig_ribbon_zigzag}(a), the lower edge is entirely composed of A-type carbon atoms, while the upper edge is made up of B-type carbon atoms. Therefore, the boundary conditions can be imposed on the two sublattices separately. In particular, the wavefunction of the A-type sublattice is required to vanish on the upper edge and the wavefunction of the B-type sublattice is required to vanish on the lower edge. With these restrictions, the eigenfunctions of the Dirac equation can be separated in two groups. The first group includes states with wavenumber $k>1/W$, where $W$ is the width of the ribbon. They are surface states and decay exponentially from the edges of the ribbon as exp($-kx$). The second group of eigenfunctions with $k<1/W$ corresponds to confined states with nodes along the transverse section of the ribbon. To obtain the exact shape of these bands within the simple tight-binding approximation, a different approach can be used, viewing the GNR as a macromolecule with a basis of four carbon atoms~\cite{MalyshevaOnipkoGNR}.

In this case too, further information can be obtained by expressing the tight-binding Hamiltonian on the Bloch sums. It is easy to see that the $k$-dependent Hamiltonian is a tridiagonal $2N\times 2N$ matrix with alternating off-diagonal elements \cite{CRE_PRB77}
\begin{equation} \label{eq:HK}
\mathbf{H(k)}=\gamma_0\left(\begin{array}{cccccc} 
	0&2 {\rm c}(k)&0&...&...&0\\[3mm]
	2{\rm c}(k)&0&1&0&...&0\\[3mm]
  0&1&0&2{\rm c}(k)&0&...\\[3mm]
  ...&0&2{\rm c}(k)&0&...&...\\[3mm]
  ...&...&...&...&...&2{\rm c}(k)\\[3mm]
  0&...&...&0&2{\rm c}(k)&0
\end{array}\right)
\end{equation}
where $c(k)=\cos(ka/2)$ and the Bloch sums have been ordered as $A_1,B_1,A_2,...,A_N,B_N$. This Hamiltonian is equivalent to the Hamiltonian of the $2N$-orbitals chain reported in Fig. \ref{fig_ribbon_zigzag}(b). By diagonalizing matrix (\ref{eq:HK}) numerically, the band structure is obtained for $N=32$, see Fig. \ref{fig_ribbon_zigzag}(c). The energy bands present some typical features: (i) the six Dirac points of the 2D graphene are mapped into $k=\pm 2\pi/(3a)$, (ii) there are two partially flat degenerate bands with 0 energy between the Dirac points and the border of the Brillouin zone, the corresponding states are mainly located at the edges, (iii) the bands are highly degenerate at the borders of the Brillouin zone, (iv) at the Dirac points and close to the charge neutrality point the levels are equispaced. All these characteristics can be understood by simple considerations on the equivalent chain of Fig. \ref{fig_ribbon_zigzag}(b). For example, at the edges of the first Brillouin zone we have $c(k=\pm \pi/a)=0$, thus the $2N$-orbitals chain is split into two isolated orbitals at its end and $N-1$ couples of dimers. This configuration generates two eigenstates with zero energy and located exactly on the edges of the ribbon, and $N-1$-fold degenerate eigenvalues $E=\pm\gamma_0$, as observed at point (iii). If we consider a very large ribbon, our chain can be thought as semi-infinite. In this case, an analytical expression for the retarded Green's function projected on the first ($A_1$) site can be obtained \cite{CRE_PRB77}
\beq G(E,k)=
\displaystyle \frac{E^2-4\gamma_0^2c^2+\gamma_0^2\pm\sqrt{(E^2-4\gamma_0^2c^2+\gamma_0^2)^2-4\gamma_0^2 E^2}}{4\gamma_0^2 E} \ ,
\eeq
where the sign in front of the square root is positive if $0<c(k)\leq 1/2$ and negative if $1/2<c(k)\leq 1$. In order to have a nonvanishing density-of-states (DoS), the discriminant under the square root must be negative, since the DoS is proportional to the imaginary part of the Green's function. Therefore, the four curves $E(k)=\pm\gamma_0\pm 2\gamma_0\cos(ka/2)$, along which the discriminant is vanishing, define the boundaries of the band structure and their intersection at $k=\pm 2\pi/(3a)$ identifies the folded Dirac points. As in the case of armchair nanoribbons, the mapping of the Dirac points could be predicted by the projection of the 2D graphene bands onto the axis corresponding to the zigzag orientation. When $E\rightarrow 0$, the retarded Green's function has a continuum of poles for $0<c(k)\leq 1/2$, which correspond to the flat bands for $|k|>2\pi/(3a)$. The structure of the energy bands around the Dirac points and for finite width ribbons is obtained by considering that $c(k=\pm 2\pi/3a)=1/2$, i.e. the hopping energies of the equivalent chain become all equal to $\gamma_0$. The eigenvalues of such an Hamiltonian are known in the literature \cite{SUT} and, for large ribbons and around the CNP, can be approximated as
\beq E_n=|\gamma_0|\frac{\pi}{N+1/2}\left(n+\frac{1}{2} \right) \ \ \ \ \ \ {\rm with} \ \ \ \ \ n=0,\pm1,\pm2,... \ .\eeq
As observed in point (iv), this means that these levels are equispaced and the spacing is inversely proportional to the width of the ribbon itself.

\begin{figure*}
 \centering \includegraphics{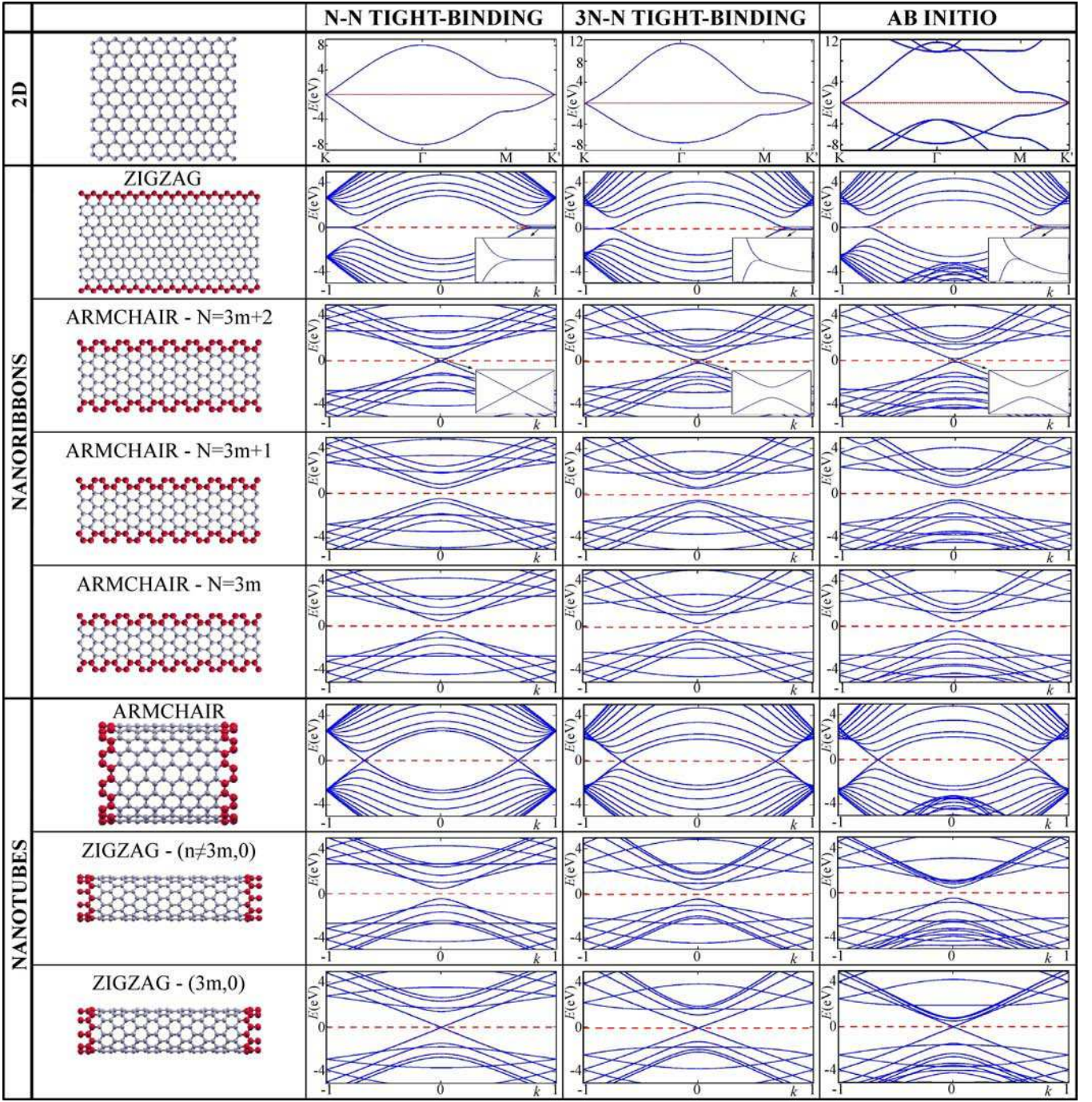}
 \caption{Band structures of 2D graphene, (9,0) and (10,0) zigzag and (10,10) armchair CNTs, and 10-zGNR, 9-10-11 aGNRs obtained by n-n tight-binding model, third n-n tight-binding model and {\it ab initio} calculations. The energy bands of 2D graphene are plotted along the K-$\Gamma$-M-K' direction. The energy bands of the GNRs and CNTs are plotted in the one-dimensional first Brillouin zone. The wavevector $k$ is in unit of $\pi$ over the lattice constant.}
 \label{mosaic}
\end{figure*}

The n-n tight-binding bandstructure of zigzag nanoribbons are summarized in Fig. \ref{mosaic} for $N=$10.

As in the case of aGNRs, the even or odd number of chains that compose the ribbons determines the mirror or glide plane symmetry of zGNRs. Even-indexed zigzag ribbons (as the 10-zGNR in Fig. \ref{mosaic}) are invariant under mirror reflection with respect to a plane perpendicular to the ribbon and containing its axis. Odd-indexed zigzag ribbons are invariant under glide plane symmetry. The parity of the wavefunctions with respect to the spatial symmetry of the GNR is clearly visible in Figs. \ref{fig_ribbon_zigzag}(d,e), where we show the 2D-projection at the $\Gamma$ point of the two wavefunctions just below and above the Fermi level for perfect asymmetric (d) and symmetric (e) zigzag ribbons (calculated with the SIESTA package \cite{SAN_IJQC65}). The spatial symmetry of zGNRs and the respective parity of the wavefunctions play an important role in determining the electronic transport properties in the presence of external fields \cite{RYC_NP3,AKH_PRB77,LI_PRL100, Alessandro}. For example, a valley-valve effect (blocking of the electrical current by a \textsl{p-n} junction) has been demonstrated theoretically in zigzag ribbons in the presence of an external potential \cite{RYC_NP3}. This effect depends critically on the parity of the number of zigzag chains across the ribbons \cite{AKH_PRB77} and has been successfully explained by realizing that, in case of even-indexed zGNRs, the symmetry of the ribbons is preserved by the considered superimposed potential, thus forbidding potential-induced transitions between opposite parity eigenstates \cite{Alessandro}. A similar explanation has been found for the case of an applied bias in zigzag ribbons, when transitions between the valence and the conduction bands are only allowed for asymmetric ribbons \cite{LI_PRL100}, and the opening of a conductance gap in the vicinity of the Fermi level is expected for symmetric zigzag ribbons. This intriguing particularity has no analogue in nanotubes.

\subsubsection{Carbon nanotube band structure}
As for graphene nanoribbons, the electronic properties of nanotubes are strongly modulated by small structural variations, in particular, their metallic or semiconducting character is determined by the diameter and helicity (chirality) of the carbon atoms in the tube. This dependence is easily understood starting from the energy dispersion relations of graphene around the Fermi level, by retaining only the $\pi$-$\pi^{*}$ bands. Consequently, rolling the graphene into a cylinder imposes periodic boundary conditions for the wavefunctions along the circumference of the tube and results in a quantization of the momentum component along this direction. For derivation details of the tight-binding graphene dispersion relations, and the subsequent zone-folding procedure to obtain the band-structure of nanotubes, we refer the reader to Ref.~\cite{RMP}. The obtained bands for armchair tubes with helicity $(N,N)$ are, for instance, given by
\begin{equation}
   E_{q}^{\pm}(k) = {\displaystyle \pm \gamma_{0} \sqrt{1\pm 4\cos\frac{ka}{2} \cos\frac{q\pi}{N}+ 4\cos^{2} \frac{ka}{2}}} \ ,
\end{equation}
where $q$ (= $1,2,...,2N$) specifies the discrete part of the wavevector perpendicular to the tube axis (i.e., the band index), $k$ is the continuous component that describes eigenstates in a given sub-band ($-\pi < ka < \pi$) and $a = 2.46\Angs$ is the graphene lattice constant.

The bandstructure for the armchair (10,10) and the zigzag (9,0) and (10,0) nanotubes calculated within the n-n tight-binding model are shown in Fig. \ref{mosaic}.

\subsection{\label{abinitio}\textsl{Ab initio} results}
In spite of the good description the most simplified n-n $\pi$-orbital tight-binding model provides for the electronic properties of 2D-graphene and carbon nanotubes, some particular features regarding mainly the effect of edges in graphene nanoribbons have only been unveiled by means of \textsl{ab initio} simulations \cite{MIY_PRB59,KAW_PRB62,Louie,Barone,Son_gaps,White_gaps} or more sophisticated tight-binding models \cite{Fujita_gaps,Sasaki,White_gapsTB}. The stability of edge states in zigzag GNRs has been analyzed both with and without H atoms by means of \textsl{ab initio} simulations. The almost flat bands in the vicinity of the Fermi level, originated by the presence of very localized states at edge atoms of the ribbon, become more dispersive, thus producing a peak in the conductance of the system, that jumps from one to three conduction channels in a region of $\approx$ 0.3 eV below the CNP of the ideal ribbon. The dispersion at the edge state predicted by first-principles calculations has been attributed to the interaction with next-nearest neighbors, that decreases the energy eigenvalue of the edge state \cite{Sasaki}, though there are still some discrepancies with respect to experimental data \cite{KLU_ASS161,NII_ASF241,KOB_PRB71}. However, the metallic character of zigzag ribbons is preserved as far as the spin degree of freedom is not taken into account \cite{Louie}.

On the other hand, \textsl{ab initio} studies have demonstrated that there are no truly metallic armchair graphene nanoribbons (\cite{Barone,Son_gaps,White_gaps}). Even for those armchair ribbons predicted to be metallic by the n-n tight-binding model, a gap opens, thus modifying their character from metallic to semiconducting. The magnitude of the gap, however, decreases with increasing ribbon width, and for a ribbon of $\approx$ 5 $\Angs$ with a metallic behavior predicted by the n-n tight-binding model, the \textsl{ab initio} estimated band gap is only $\approx$ 0.05 eV. The magnitude of the gaps of the semiconducting ribbons predicted by the n-n tight-binding model also differs from those estimated by means of first-principles calculations. There have been several explanations for the origin of these gaps, supported by more sophisticated tight-binding models that take into account edge distortion \cite{Fujita_gaps, Son_gaps}, up to three nearest neighbors interactions \cite{White_gaps}, or both \cite{White_gapsTB}. 

The {\it ab initio} bandstructure of 2D graphene and different types of nanoribbons and nanotubes are shown in the last column of Fig. \ref{mosaic}. All the {\it ab initio} calculations have been performed with the SIESTA package \cite{SAN_IJQC65}.

\subsection{Third nearest neighbors interaction tight-binding model}

\subsubsection{Changes in 2D graphene band structure}
\begin{figure}[htb]
\begin{center}
   {\resizebox{8cm}{!}{\includegraphics{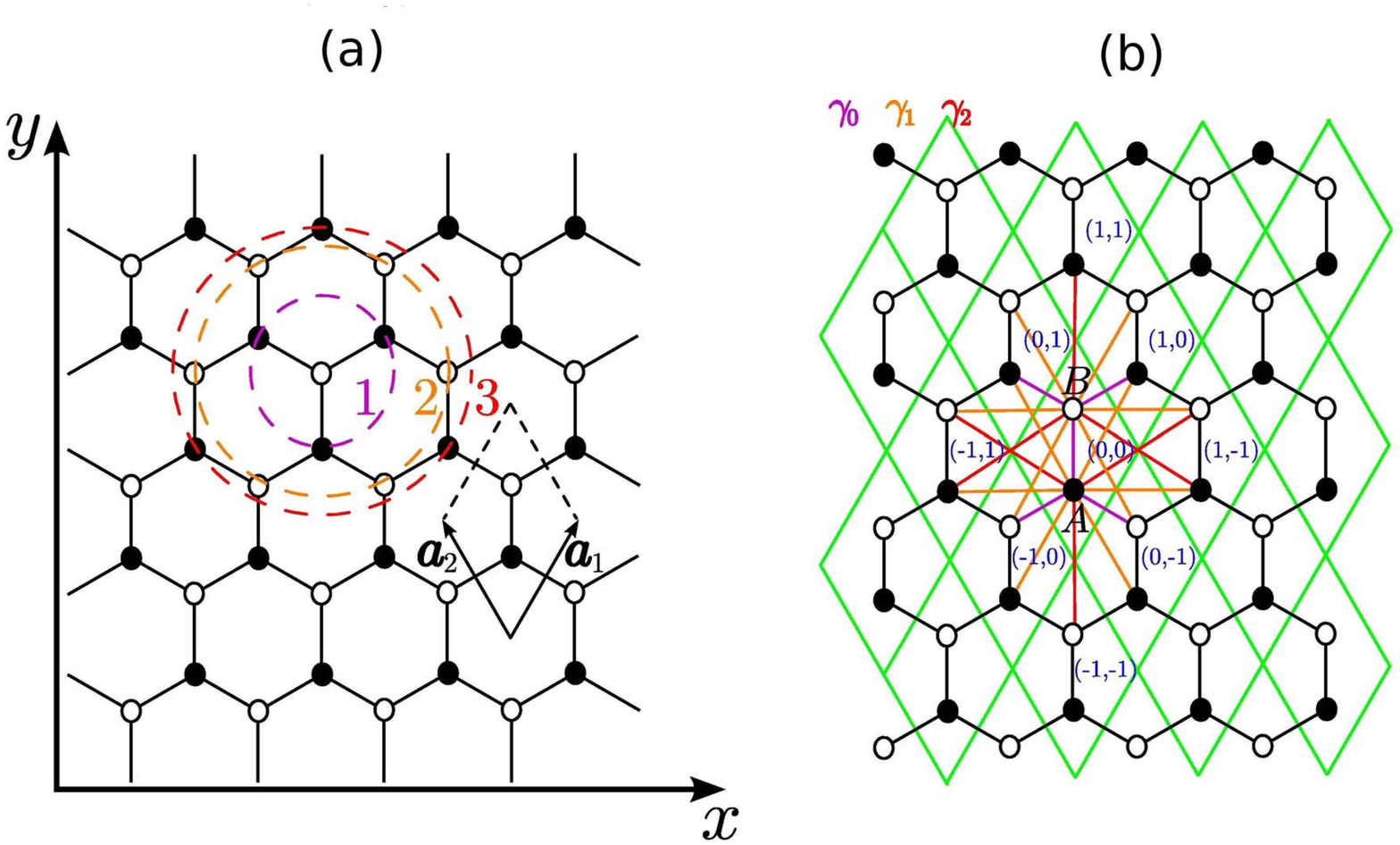}}}
  \caption{(a) An atom in the graphene lattice with its first, second and third neighbors, indicated by three dashed circles. Second neighbors belong to the same sublattice of the considered atom, first and third neighbors to opposite sublattice type. (b) A unit cell of the graphene lattice and its adjacent unit cells. Boundaries between unit cells are marked in green, they are labeled in blue. Purple lines represent first, orange lines for second and red lines for third nearest neighbor hopping.} \label{fig:graphene_neighbours}
\end{center}
\end{figure}

In section \ref{section_2D}, it was mentioned that the band structure of 2D graphene can be calculated using the nearest-neighbor tight-binding approximation \cite{Wallace1947}. This leads to Eq.\ (\ref{eq:2D_bands_1st_NN}), which is plotted in Fig.\ \ref{FIG2} (central panel) and Fig.\ \ref{mosaic}. We will now see how this result is modified by taking into account higher orders of interaction and orbital overlap.

Figure \ref{fig:graphene_neighbours}(a) graphically shows the distances between first, second and third nearest neighbors. These distances are given by $d_1=1.42\Angs$, $d_2=\sqrt{3}d_1\approx1.73d_1$ and $d_3=2d_1$. From Fig.\ \ref{fig:graphene_neighbours}(a), we also see that every atom in the graphene lattice has three nearest neighbors, six second-nearest neighbors and three third-nearest neighbors. The small difference in length between the distances of second and third nearest neighbors ($d_2$ and $d_3$) suggests the inclusion of both of these interactions.
Figure \ref{fig:graphene_neighbours}(b) allows us to evaluate which interactions exist between the unit cell here labeled (0,0) and its neighboring unit cells. We can therefore write the tight-binding Hamiltonian, including up-to-third nearest neighbor interaction:
\begin{eqnarray}
      \mathbf{H}(\kvec) &=& \mathbf{H}_{0,0}+e^{\I\kvec\cdot\aone}\mathbf{H}_{1,0}+e^{-\I\kvec\cdot\aone}\mathbf{H}_{-1,0} 
      +e^{\I\kvec\cdot\atwo}\mathbf{H}_{0,1}+e^{-\I\kvec\cdot\atwo}\mathbf{H}_{0,-1} 
      +e^{\I\kvec\cdot(\aone+\atwo)}\mathbf{H}_{1,1}+e^{-\I\kvec\cdot(\aone+\atwo)}\mathbf{H}_{-1,-1} \notag\\
               &+& e^{\I\kvec\cdot(\aone-\atwo)}\mathbf{H}_{1,-1}+e^{-\I\kvec\cdot(\aone-\atwo)}\mathbf{H}_{-1,1} \label{eq:hextmatrix}
\end{eqnarray}
where $\aone=(\sqrt{3},3)a/2$ and $\atwo=(\sqrt{3},-3)a/2$ are the translation vectors of the hexagonal lattice and
\beq
\begin{array}{lllllllll}
  \mathbf{H}_{0,0}  &=& \begin{pmatrix} \epsilon_0 & -\gamma_0 \\ -\gamma_0 & \epsilon_0 \end{pmatrix} &
  \ \ \ \ \mathbf{H}_{1,-1} &=& \begin{pmatrix} -\gamma_1 & -\gamma_2 \\ -\gamma_2 & -\gamma_1 \end{pmatrix} & 
  \ \ \ \ \mathbf{H}_{0,1}  &=& \begin{pmatrix} -\gamma_1 & -\gamma_0 \\ 0 & -\gamma_1 \end{pmatrix} \\[5mm]
  \mathbf{H}_{1,1}  &=& \begin{pmatrix} 0 & -\gamma_2 \\ 0 & 0 \end{pmatrix} &
  \ \ \ \ \mathbf{H}_{0,-1} &=& \begin{pmatrix} -\gamma_1 & 0 \\ -\gamma_0 & -\gamma_1 \end{pmatrix} &
  \ \ \ \ \mathbf{H}_{-1,-1}&=& \begin{pmatrix} 0 & 0 \\ -\gamma_2 & 0 \end{pmatrix} \ ,
\end{array}\eeq 
with $\mathbf{H}_{i,j}=\mathbf{H}_{j,i}$ for $i\neq j$. First, second and third nearest neighbor hopping constants are here called $\gamma_0$,$\gamma_1$ and $\gamma_2$, in accordance with Fig.\ \ref{fig:graphene_neighbours}(b). The onsite energy is indicated by $\epsilon_0$. Analogously, we define $\sigma_0$, $\sigma_1$ and $\sigma_2$ to be the overlap matrix elements between first, second and third nearest neighbors, respectively, and we get a formula for $S$ that looks just like Eq. (\ref{eq:hextmatrix}), with every $\mathbf{H}$ replaced by an $\mathbf{S}$ and
\beq
\begin{array}{lllllllll}
  \mathbf{S}_{0,0}  &=& \begin{pmatrix} 1 & \sigma_0 \\ \sigma_0 & 1 \end{pmatrix} &
  \ \ \ \ \mathbf{S}_{1,-1} &=& \begin{pmatrix} \sigma_1 & \sigma_2 \\ \sigma_2 & \sigma_1 \end{pmatrix} &
  \ \ \ \ \mathbf{S}_{0,1}  &=& \begin{pmatrix} \sigma_1 & \sigma_0 \\ 0 & \sigma_1 \end{pmatrix} \\[5mm]
  \mathbf{S}_{1,1}  &=& \begin{pmatrix} 0 & \sigma_2 \\ 0 & 0 \end{pmatrix} &
  \ \ \ \ \mathbf{S}_{0,-1} &=& \begin{pmatrix} \sigma_1 & 0 \\ \sigma_0 & \sigma_1 \end{pmatrix} &
  \ \ \ \ \mathbf{S}_{-1,-1}&=& \begin{pmatrix} 0 & 0 \\ \sigma_2 & 0 \end{pmatrix} \ ,
\end{array}\eeq
where of course $\mathbf{S}_{i,j}=\mathbf{S}_{j,i}$ for $i\neq j$. We now define
\beq\begin{array}{ll}
  g_0(\kvec) &= 1+e^{\I\kvec\cdot\aone}+e^{\I\kvec\cdot\atwo} \\[3mm]
  g_1(\kvec) &= e^{\I\kvec\cdot\aone}+e^{-\I\kvec\cdot\aone}+e^{\I\kvec\cdot\atwo}+e^{-\I\kvec\cdot\atwo}
                 +  e^{\I\kvec\cdot(\aone-\atwo)}+e^{-\I\kvec\cdot(\aone-\atwo)} \\[3mm]
  g_2(\kvec) &= e^{\I\kvec\cdot(\aone+\atwo)}+e^{\I\kvec\cdot(\aone-\atwo)}+e^{-\I\kvec\cdot(\aone-\atwo)} \ ,
\end{array}\eeq
which permits us to write the Hamiltonian $\mathbf{H}$ as
\begin{equation}
  \mathbf{H} = \begin{pmatrix} \epsilon_0 - \gamma_1 g_1(\kvec) & -\gamma_0 g_0(\kvec)-\gamma_2 g_2(\kvec) \\
                          -\gamma_0 g^*_0(\kvec)-\gamma_2 g^*_2(\kvec) & \epsilon_0 - \gamma_1 g_1(\kvec) \end{pmatrix}
  \label{eq:GN_Hamiltonian}
\end{equation}
and the overlap matrix $\mathbf{S}$ as
\begin{equation}
  \mathbf{S} = \begin{pmatrix} 1+\sigma_1 g_1(\kvec) & \sigma_0 g_0(\kvec) + \sigma_2 g_2(\kvec) \\
                          \sigma_0 g^*_0(\kvec) + \sigma_2 g^*_2(\kvec) & 1+\sigma_1 g_1(\kvec) \end{pmatrix}.
  \label{eq:GN_overlap_matrix}
\end{equation}

Now all that is left to do is to solve the generalized characteristic equation
\beq\begin{array}{r}
  |\mathbf{H}-E\mathbf{S}| = \begin{vmatrix} H_{AA}-ES_{AA} & H_{AB}-ES_{AB} \\ H_{AB}^*-ES_{AB}^* & H_{AA}-ES_{AA} \end{vmatrix} = 0 
  \Rightarrow (H_{AA}-ES_{AA})^2 - (H_{AB}-ES_{AB})(H_{AB}^*-ES_{AB}^*) = 0 \\[3ex]
  \Rightarrow E^2(\overbrace{S_{AA}^2 - |S_{AB}|^2}^{\epsilon_1}) + E(\overbrace{H_{AB}S_{AB}^*+H_{AB}^*S_{AB}}^{\epsilon_2} - 2\overbrace{H_{AA}S_{AA}}^{\epsilon_3}) + \overbrace{H_{AA}^2 - |H_{AB}|^2}^{\epsilon_4} = 0 \ ,
\end{array}\eeq
where we introduced $\epsilon_{1,2,3,4}$ as a shorthand notation. From comparison between Eqs.(\ref{eq:GN_Hamiltonian}) and (\ref{eq:GN_overlap_matrix}) we see that
\beq\begin{array}{lcl}
  \epsilon_1 &=& \left(1+\sigma_1 g_1(\kvec)\right)^2 - \left|\sigma_0 g_0(\kvec) + \sigma_2 g_2(\kvec)\right|^2 \\[3mm]
  \epsilon_2 &=& -\left[\left(\gamma_0 g_0(\kvec)+\gamma_2 g_2(\kvec)\right)\times\left(\sigma_0 g^*_0(\kvec) + \sigma_2 g^*_2(\kvec)\right) + c.c. \right] \\[3mm]
  \epsilon_3 &=& \left(\epsilon_0-\gamma_1 g_1(\kvec)\right)\times\left(1+\sigma_1 g_1(\kvec)\right) \\[3mm]
  \epsilon_4 &=& \left(\epsilon_0-\gamma_1 g_1(\kvec)\right)^2 - \left|\gamma_0 g_0(\kvec)+\gamma_2 g_2(\kvec)\right|^2 \ .
\end{array}\eeq

We are thus left with the following dispersion relation for 2D graphene:
\begin{equation}
    E^\pm = \frac{-(\epsilon_2-2\epsilon_3)\pm\sqrt{(\epsilon_2-2\epsilon_3)^2-4\epsilon_1\epsilon_4}}{2\epsilon_1}. \label{eq:quadeq}
\end{equation}
The values for $\gamma_{0-2}$ and $\sigma_{0-2}$ have been calculated by Reich et al.\ by comparing and fitting band structures from {\it ab initio} and tight-binding calculations \cite{Reich2002}. The third-nearest-neighbor tight-binding band structure with these parameters is plotted in Fig.\ \ref{mosaic}. Recently, more sophisticated GW {\it ab initio} calculations \cite{GRU} for bilayer and few-layer graphene provided $\sim 20\%$ larger values of $\gamma$s. This is in better agreement with some experimental measurements, in particular angle-resolved photoemission (ARPES), where the long-range Coulomb interaction plays an important role and the band structure at high energies requires an more accurate description.

\subsubsection{Effects on the band structure of GNRs}
It was already mentioned that the metallicity of armchair GNRs of widths $N=3m+2$ ($m\in \mathbb{N}$) is only found in the n-n tight-binding approximation, while \textit{ab initio} calculations show all armchair GNRs to be semiconducting (cf. section \ref{section_nnTB} and section \ref{abinitio}). This finding is reproduced by third-nearest-neighbor tight-binding calculations.
\begin{figure}[htb]
  \centering
  {\resizebox{8cm}{!}{\includegraphics{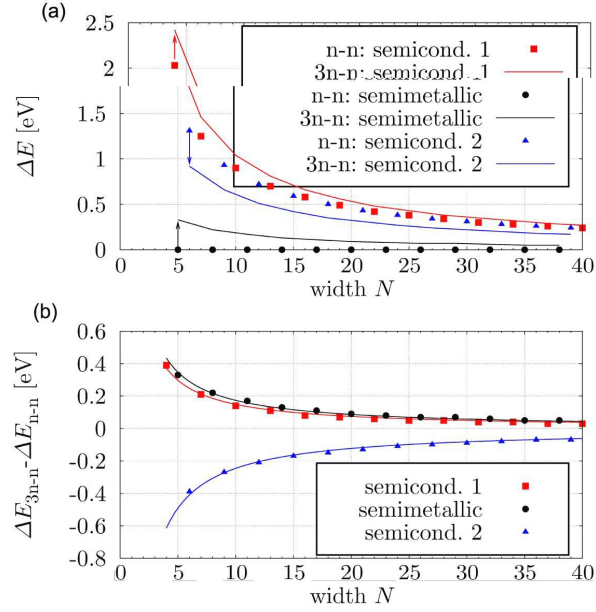}}}
  \caption{(a) The band gaps of armchair GNRs as a function of their width $N$ in n-n and 3n-n tight-binding. Because of the three-fold periodicity in the band gap's width, it is convenient to plot three different curves or sets of points for $N \bmod 3 = 0,1,2$. The symbols show n-n, the lines corresponding 3n-n results. (b) The gap opening/closing obtained from a comparison of n-n and 3n-n tight-binding results. The symbols show the 
opening (or closing) of the band gap in armchair GNRs of the respective widths. The lines are fit to $\alpha/N$, where $\alpha$ is the fitting parameter. This is analog to the case of zigzag carbon nanotubes in which the gap decreases as $1/d$, where $d$ is the diameter of the tube.}
  \label{fig:GN_band_gaps}
\end{figure}
The gap opening can be quantified by comparing n-n and 3n-n tight-binding results with each other (cf. Fig. \ref{fig:GN_band_gaps}). The relative opening or closing of the band gap decreases with the width as $1/N$. Furthermore, it has been shown that edge deformations \cite{ZHE_PRB75, WAN_PRB75} and strain \cite{FIO_IEEE, SUN_ARXIV} significantly change the band gap of armchair GNRs, turning metallic ribbons into semiconducting ones, even when calculated with the n-n tight-binding model. The metallicity of certain armchair ribbons should therefore be considered an artifact of the n-n tight-binding approximation and not a robust feature of these ribbons.

In the case of zigzag GNRs the 3n-n tight-binding model shows robust metallicity and a higher dispersion of the edge state than in the case of only nearest-neighbor interaction (cf. Fig. \ref{mosaic}). The full \textit{ab initio} results show a band gap that opens due to magnetic ordering \cite{Son_gaps}. This feature is not reproduced by the tight-binding model employed here, but one can account for such spin-related effects by including a Hubbard term in the Hamiltonian \cite{FUJ_JPSJ65,GUN_NL7,PAL_PRB77, FER_PRB77}.

It can thus be said that the third-nearest-neighbor tight-binding method leads to significantly better results than the nearest-neighbor method when compared to \textit{ab initio} results, while keeping within the simplicity of the tight-binding model.

The 3n-n energy bands for 2D graphene and different types of nanoribbons and nanotubes are summarized in the fourth column of Fig. \ref{mosaic}.

\subsection{Transport in nanotubes and nanoribbons}
When neglecting disorder effects, a carbon nanotube of length $L$ in between metallic contact reservoirs is a ballistic conductor with a $L$-independent conductance given by the energy-dependent number of available quantum channels $N_{\perp}(E)$. Each channel carries one conductance quantum $G_{0}=2e^{2}/h$, thus $G(E)=2e^{2}/h\times N_\perp(E)$, including spin degeneracy. This only occurs in case of perfect (reflection-less) or ohmic contacts between the CNT and metallic voltage probes~\cite{NEM_PRL96,NEM_PRB77}. In this regime, the expected energy-dependent conductance spectrum is easily deduced from band structure calculations, by counting the number of channels at a given energy. For instance, armchair nanotubes present two quantum channels at the charge neutrality point, which result in $G(E_F)=2G_0$. At higher energies, the ballistic conductance increases as more channels become available to conduction.

In contrast, the eigenstates of defect-free armchair GNRs in the first plateau are constrained by the hard-wall (Dirichlet) boundary conditions which require a node at the edges, thus excluding the cosine solutions \cite{ARE_NL7}. As a result, a singly degenerate band is left in the first conductance plateau.

Within the linear regime near the CNP, a closed expression for the the number of channels is given by
\beq
N_{\perp}(E) = 2 + 4 \left( \frac{d}{2 \hbar v_\mathrm{F}} \left|E-E_\mathrm{F}\right| \right)
\eeq
in the case of an armchair CNT. For CNTs of other chiralities and GNRs, similar expressions can be obtained.
	
\subsubsection{\label{pseudo}"Pseudodiffusive" transport in clean graphene nanoribbons}
\begin{figure}[htp]
  \centering
  \includegraphics[width=8cm]{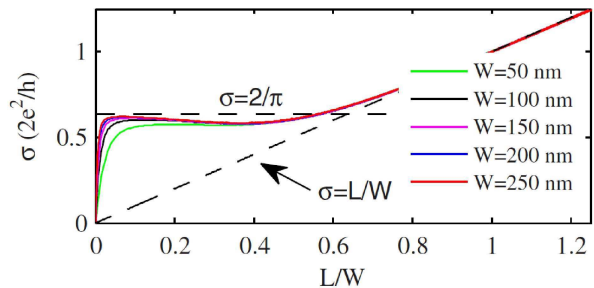}
  \caption{Conductivity $\sigma$ of a zigzag ribbon as a function of its aspect ratio (length $L$ over width $W$), in the case of graphene leads and energy just above the charge neutrality point. For short and wide nanoribbons ($0.1<L/W<0.4$), the value of the conductivity is very close to $4e^2/\pi h$. This figure is taken from [Cresti, A.; Grosso, G.; Pastori Parravicini, G. {\it Phys. Rev. B} {\bf 2007}, 76, 205433].}
  \label{fig_pseudo}
\end{figure}

Transport through short and wide clean graphene monolayers around the charge neutrality point shows very peculiar behaviors analogous to those observed in diffusive coherent disordered conductors. This turns out to be a mere but extremely surprising coincidence, due to the particular distribution of the transmission coefficients. In two first theoretical studies, Tworzyd{\l}o et al. \cite{TWO_PRL96} and Katsnelson \cite{KAT_EPJB51}  have investigated the transmission coefficients for a ribbon of length $L$ and width $W$ in the limit of short and wide armchair ribbons ($L<W$) with hardwall or smooth confining potential at the edges. The system is kept at energies around the neutrality point, and it is connected to two leads at high potential with a large number $N$ of active conductive channels. The electronic transmission is thus mainly sustained by tunneling through evanescent modes, and an analysis based on the Dirac equation leads to
\beq T_n\approx \frac{1}{{\rm cosh}^2[\pi(n+1/2)L/W]} \ \ {\rm with} \ \ n=0,\pm1,\pm2,...\eeq     
for the hardwall confining potential, and a very similar expression for the smooth potential \cite{TWO_PRL96}. As a consequence, in the limit of a very large number of transmission channels and high aspect ratio ($N>>W/L\rightarrow\infty$)
\beq G=\frac{4e^2}{h}\sum_n T_n =\frac{4e^2}{\pi h} \frac{W}{L} \ \ \ \ ;  \ \ \ \ F=\frac{\sum_n T_n(1-T_n)}{\sum_n T_n}=\frac{1}{3}\ ,\eeq
where $G$ is the conductance and $F$ is the Fano factor, i.e. the ratio between shot noise and Poissonian noise. In this limit, the system is ohmic with minimum conductivity $\sigma_{\rm min}=4e^2/\pi h$ and Fano factor equal to 1/3, as in the case of a diffusive conductor. This theoretical result has been tested numerically by calculations based on a tight-binding model \cite{TWO_PRL96,CRE_PRB76}, and a good agreement was found, see Fig. \ref{fig_pseudo}.

Schomerus has recently proved \cite{SCH_PRB76} that this behavior is essentially independent of the specific configuration of the contacts, provided that the graphene system is very close to the neutrality point and the leads support a large number of propagation modes. The introduction of disorder is expected to affect the Fano factor, but the literature does not agree on the results. In one case \cite{SAN_PRB76} the Fano factor turns out to be lowered by disorder, but still keeps a value close to 0.3. In another case \cite{LEW_PRB77}, the Fano factor is found to be a little lower than 1/3 for very weak disorder with a small peak at the neutrality point, and then to stay around 1/3 for short and wide ribbons for weak disorder, and to go over 1/3 for stronger disorder.

Three recent experimental works \cite{MIA_SCI317,DIC_PRL100,DAN_PRL100} on exfoliated graphene samples confirm the main theoretical predictions. Miao and co-workers \cite{MIA_SCI317} have first evidenced that for wide and short graphene strips ($W/L>4$) the minimum measured conductivity approaches the value $4e^2/\pi h$. This result has been observed in particular for very broad ribbons ($W>>L$), because in this case boundary effects are negligible. DiCarlo and co-workers \cite{DIC_PRL100} and Danneau and co-workers \cite{DAN_PRL100} performed measurements of the shot noise and observed the predicted value $F=1/3$ for clean samples. Moreover, some other predictions \cite{LEW_PRB77} seems to be confirmed, i.e. the increasing value of the Fano factor for weakly disordered systems \cite{DAN_PRL100} and the presence of a (larger than expected) peak at the neutrality point for very clean ribbons \cite{DIC_PRL100}.

The coincidence between the behaviors of coherent disordered conductors and short and wide ballistic graphene ribbons is a puzzling issue, whose origin, if any, is not clearly understood.

\subsubsection{High magnetic fields and spatial chirality of currents in GNRs}
\begin{figure}[ht]
  \centering
  \includegraphics[width=8cm]{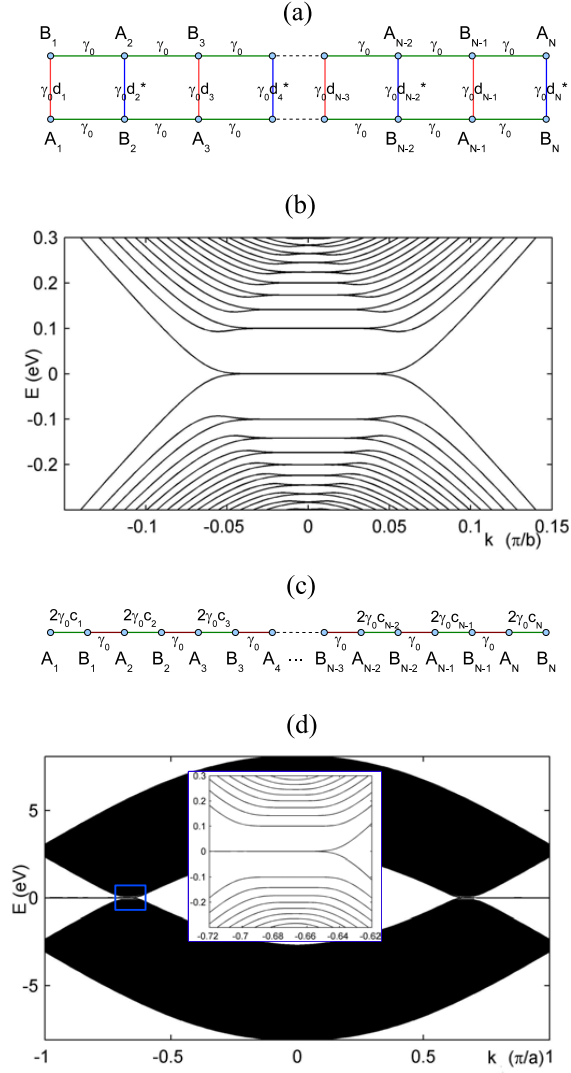}
  \caption{\label{fig_ribbon_magnetic}(a) Coupled chains equivalent to an armchair ribbon threaded by a perpendicular and homogeneous magnetic field $B$. (b) Energy bands of an aGNR with $N=611$, corresponding to a width of about 75 nm, for $B=10$ T. (c) Chain equivalent to a zigzag ribbon threaded by a perpendicular and homogeneous magnetic field $B$. (d) Energy bands of a zGNR with $N=353$, corresponding to a width of about 75 nm, for $B=10$ T.}
\end{figure}
As seen in section \ref{section_2D}, a homogeneous perpendicular magnetic field $B$ that threads a 2D graphene sheet induces Landau levels spaced as $\sqrt{n}$ and proportional to $\sqrt{B}$ \cite{MCC_PR104,Nemec_PRB_2007}. A prescription for the inclusion of magnetic field effects into the band structure of carbon-based nano-networks can be found in Ref. \cite{Nemec_PRB_2006}. An energy structure similar to that of 2D graphene is also observed for bulk electrons in nanoribbons, provided that the width $W$ of the ribbon is larger than the magnetic length. At the borders of the Brillouin zone, the bands bend upward or downward due to the finite width of the ribbon. The corresponding magnetic states are located at the upper and lower edges.

In the simple n-n tight-binding model described in section \ref{section_nnTB}, we can account for the magnetic field by means of the Peierls phase factors on the hopping parameters. By choosing the gauge properly, the Hamiltonian preserves the translation invariance along the longitudinal axis of the ribbon and can be conveniently expressed on the Bloch sums basis. Armchair and zigzag ribbons are still equivalent to a $2N$-orbitals chain and two coupled $N$ orbital chains, respectively, but the hopping energies now depend on the chain index $n=1,2,...,N$ \cite{CRE_PRB77}. In the case of aGNRs, the coupling parameters between the two chains become (see Fig. \ref{fig_ribbon_magnetic}(a)) 
\beq d_n(k)= \exp\left[\pm \left(\frac{kb}{2}-\frac{n\alpha}{2}+\frac{N+1}{4}\alpha\right)\right] \ ,\eeq
where $\alpha=2\pi\Phi(B)/\Phi_0$ is proportional the ratio between the magnetic flux $\Phi$ through a hexagonal plaquette of the honeycomb lattice and the elementary magnetic flux $\Phi_0$. In the case of zGNRs, the hopping $k$-dependent parameters between the orbitals of the chain become (see Fig. \ref{fig_ribbon_magnetic}(c)) 
\beq c_n(k)=\cos\left(\frac{ka}{2}-\frac{n\alpha}{2}+\frac{N+1}{4}\alpha \right) \ . \eeq
A direct numerical diagonalization of the Hamiltonian provides the energy bands for a $N=611$ armchair ribbon (Fig. \ref{fig_ribbon_magnetic}(b)) and $N=353$ zigzag ribbon (Fig. \ref{fig_ribbon_magnetic}(d)) for $B=10$ T. We consider ribbons much wider (about 75 nm) than those considered in the absence of magnetic field, in order to have the onset of Landau levels at relatively small magnetic fields.

In the case of armchair ribbons, see Fig. \ref{fig_ribbon_magnetic}(b), we can clearly observe the sequence of double degenerate Landau levels. At the borders of the Brillouin zone, the average transverse position of the states approaches the edges of the ribbon and the energy bands bend upward for electron-like particles and downward for hole-like particles. In fact, the particles cannot complete the cyclotron orbit due to the edges and as a consequence their energy rises. As in ordinary two-dimensional electron gas, the states on the right side of the Brillouin zone are located on the upper edge of the ribbon, while the states on the left side are located on the lower edge. The double degeneracy of the bands is removed for edge states. This can be explained in terms of the Dirac equation since the boundary conditions for armchair ribbons entail the admixing of the valleys \cite{BRE_PRB73}. As a consequence, the corresponding wavefunctions hybridize thus giving rise to the observed structure.

In the case of zigzag ribbons, see Fig. \ref{fig_ribbon_magnetic}(d), we observe a positive and a negative set of Landau levels around each of the two valleys, and a zero energy level that extends from one valley to the other. The states on the right side of each valley are located on the upper edge of the ribbon, while the states on the left side of each valley are located on the lower edge. Again, when the average transverse position of the states approaches the edges of the ribbon, the energy bands bend upward for electron-like particles and downward for hole-like particles. The structure of the energy bands around the two valleys is exactly the same. Again, this can be understood by considering the Dirac equation. In contrast to armchair ribbons, the boundary conditions for the wavefunction do not mix the two valleys and therefore, close to the charge neutrality point, they behave independently of each other. The Dirac equation also explains the peculiar structure of the two lowest Landau levels. The corresponding wavefunctions are extremely localized at one edge for a sublattice and behave as ordinary magnetic states for the other.

From the transport perspective, the first important consequence of the magnetic electronic structure of GNRs is the theoretical prediction and the experimental observation of the so called anomalous (of half integer) Hall effect \cite{NovoselovEtAl2005,ZHA_N438,PER_PRB73,GUS95}, i.e. the magnetoconductivity is given by
\beq G=\frac{4e^2}{h}\left(n+\frac{1}{2} \right) \ .\eeq
The large separation between the first Landau levels allows the observation of this phenomenon at room temperature. 

In addition to this, magnetic edge states have peculiar and interesting properties. Since the direction of the carrier group velocity depends on the sign of their energy band slope with respect to the wavevector $k$, electron-like particles can only flow from the left side of the ribbon to the right side along the upper edge, and from the right side to the left side along the lower edge. In the case of hole-like particles the energy bands are reversed and they can only flow from the left side of the ribbon to the right side along the lower edge, and from the right side to the left side along the upper edge. The spatial separation between the conductive channels that support current in opposite directions is called spatial chirality. In the case of graphene nanoribbons, electron- and hole-like particles have opposite spatial chiralities \cite{CRE_PRB77}. This property may be exploited to manipulate the current and obtain a current switch \cite{CRE_NT15}. In fact, the edge the transport currents flow along is determined by the position of the electrons energy with respect to the charge neutrality point, and their energy can be varied by means of top or back gates, even locally.     

\section{\label{section_disorder}Disorder effects}
Transport properties in graphene-based materials also turn out to be strongly affected by 
disorder, which can originate from impurities (charges trapped in the oxide, chemical impurities, etc..), topological defects (vacancies, edge disorder,...), or long range deformation modes (ripples) in 2D graphene. Importantly, the dominant scattering processes and resulting transport features are very dependent on the range of the disorder potential and the robustness or destruction of the underlying sublattice symmetries. 

For massless Dirac fermions, a long range scattering potential, i.e. with Fourier components $V({\bf q})$ such that ${\bf q} \ll {\bf K}$, will strongly reduced the intervalley scattering probability between the two non-equivalent Dirac nodes (${\bf K}_{-} \to {\bf K}_{+}$). In the one-dimensional case provided by armchair CNTs, this results in a full suppression of backscattering probability as demonstrated by Ando and co-workers \cite{AndoSaito,Whitetodorov}.  The impact of long range disorder in two-dimensional graphene remains more debated, with opened issues concerning how weak or strong localization regimes are genuinely affected by specific Dirac fermions properties 
\cite{MCC_PRL97,OST_PRB74,MOR_PRL97,KHV_PRB75}.

In contrast, the presence of short range disorder (${\bf q}\sim {\bf K}$) will allow for all possibilities of intravalley and intervalley scattering events between ${\bf K}_{+}$ and ${\bf K}_{-}$, leading to stronger backscattering and localization effects, although a true Anderson localization in two-dimensional graphene remains fiercely debated theoretically, because of the "Dirac nature" of low-energy excitations \cite{SUZ_JPSJ75,ALE_PRL97,NOM_PRL99}.

Disorder effects in the quantum coherent regime can yield localization regimes \cite{AND_PR124,ABR_PRL42,LEE_PRL47,LEE_RMP57}, and indeed several experimental evidences for weak and strong Anderson localization regimes have been reported in disordered carbon nanotubes \cite{RMP,ROC_PRL87,FED_PRL94,GOM_NM4,BIE_PRL95,STR_SST21,FED_NL7,LAS_PRL98,STO_NJP9,BIE_JPCM20,FLO_JPCM20}, but one should also stress that the effects of disorder on quantum transport in graphene-based materials of lower dimensionality, such as carbon nanotubes and graphene nanoribbons, are expected to be maximized if compared to the case of 2-dimensional graphene. Indeed, low dimensionality and confinement effects yield strong modifications of band structures with the appearance of van Hove singularities (vHs), close to which energy dispersion -or wavepacket velocity- is very low. In the forthcoming sections, we will focus on short-range disorder effects by using the Anderson-type disorder potential, and we will review its effects on electronics states, mean free paths and localization phenomena.

\subsection{Density of states in weak disordered CNTs and GNRs}
The spectrum of a system may be affected by weak disorder only through small energy shifts in the energy. A constant density of states will therefore remain unaffected by weak disorder. Regions of slowly changing DOS will only show small effects. Discontinuities and van Hove singularities will, in contrast, be strongly affected and appear smeared out at the energy scale of the disorder strength~{\cite{hgle-vhsidmqwan2002,NemecGNR}}.

The {\tmabbr{DoS}} of a general quantum wire under the influence of Anderson disorder can be obtained via an algorithm based on diagrammatic perturbation theory that takes into account localization effects by including multiple scattering~{\cite{hgle-vhsidmqwan2002}}. By dropping the crossing diagrams within the noncrossing approximation ({\tmabbr{NCA}})~{\cite{abrikosov-qftmisp1965}}, we can write the self-energy $\Sigma \left( E \right)$ to all orders as a recursive expression, which can then be iterated numerically until self-consistency is reached. Though the applicability of the {\tmabbr{NCA}} is not obvious, it can be justified by comparing the contribution of various terms at low orders~{\cite{hgle-vhsidmqwan2002}}.

The original formulation of this approach is restricted to the special case of CNTs where all atoms are equivalent through symmetry and the self-energy takes the same value for all atoms. It can, however, be generalized to arbitrarily structured quantum wires using matrix notation. The self-energy $\Sigma \left( E \right)$ that accounts for disorder is a diagonal matrix obeying the recursive relation
\begin{eqnarray}
  \left[ \Sigma \left( E \right) \right]_{i, j} & = & \delta_{ij}
  \sigma_{\varepsilon_i}^2 \left[ \left( E + \mathrm{i} 0^+ - \mathcal{H}_0 -
  \Sigma \left( E \right) \right)^{- 1} \right]_{ij} .  \label{self-energy}
\end{eqnarray}
For a periodic system, the self-energy has the same periodicity as the Hamiltonian. The block-tridiagonal matrix $\left( E + \mathrm{i} 0^+ - \mathcal{H}_0 - \Sigma \left( E \right) \right)$ can therefore be inverted numerically using a highly convergent renormalization-decimation algorithm~{\cite{sancho-hcsftcobasgf1985,nemec-qticn2007}}, allowing us to go beyond the energy range near the Fermi energy, where the special band structure allows analytic inversion.

Starting with $\Sigma = 0$, each numerical iteration of this recursive relation is equivalent to one additional perturbative order. Typically, convergence is achieved after less than ten iterations, except for energies near a van Hove singularity, where hundreds of iterations may be necessary. This clearly indicates that low-order perturbation theory breaks down near band edges.

\begin{figure}[ht]
 {\resizebox{8cm}{!}{\includegraphics{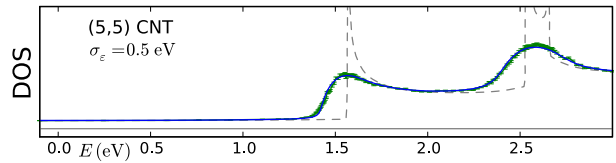}}}
 \caption{\label{disordered-DOS}Density of states $g(E)$ of an infinite armchair (5,5)~CNT under the influence of Anderson disorder. Dashed line: $g \left( E \right)$ of the clean system displaying the van Hove singularities. Solid line: $g \left( E \right)$ from Eq.~(\ref{Eqn:g(E)}). Data with error bars: values obtained numerically by sample-averaging. Adapted from [Nemec, N., Richter, K.; Cuniberti, G. {\it New J. Phys.} {\bf 2008}, 10, 065014].}
\end{figure}

\begin{figure}[ht]
 {\resizebox{8cm}{!}{\includegraphics{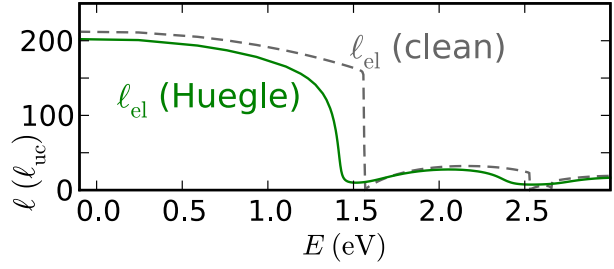}}}
 \caption{\label{l_el-Huegle}The elastic mean free path $\ell_{el}$ in a (5,5) CNT with Anderson-disorder of strength $\sigma_{\varepsilon} = 0.5 \tmop{eV}$. Both lines are obtained from Eq.~(\ref{l_el-general}). In the case of the dashed line, $g(E)$ and $N_{\tmop{ch}}$ correspond to the values of the clean system, resulting in discontinuities at the band edges, similar to Fig.~(\ref{l_el-perturbative}). For the solid line, the disorder effects on $g (E)$ and $N_{\tmop{ch}}$ are taken into account using Eqs.~(\ref{self-energy}) and (\ref{Eqn:g(E)}).}
\end{figure}

The {\tmabbr{LDoS}} of each orbital $i$ in the unit cell can now be obtained directly from the imaginary part of the Green function
\begin{eqnarray}
  g_i \left( E \right) & = & - \frac{1}{\pi} \mathrm{\tmop{Im}} \left[ \left(
  E + \mathrm{i} 0^+ - \mathcal{H}_0 - \Sigma \left( E \right) \right)^{- 1}
  \right]_{i, i} .  \label{Eqn:g(E)}
\end{eqnarray}
Figure \ref{disordered-DOS} shows this quantity in direct comparison with the numerically exact value obtained by sample averaging. The slight deviation visible at the flanks of the van~Hove singularities is caused by the {\tmabbr{NCA}}~{\cite{hgle-vhsidmqwan2002}}. The elastic mean free path $\ell_{el}$ based on the DoS of a disordered system, displayed in Fig.~\ref{l_el-Huegle}, is no longer a purely perturbative quantity, but it takes into account the scattering into localized states present at any given energy.

The DoS of aGNRs, see Fig. \ref{NN:GNRa20}(a), is obtained straightforwardly and shows the expected smoothing of van Hove singularities (including those at the gap edges) in complete analogy to the CNT. For zigzag-edge GNRs (Fig. \ref{NN:GNRa20}(b)), the most prominent feature in the DoS is the strongly broadened edge state around the Fermi energy, whose shape displays a clear deviation of the NCA from the numerical exact calculation.

\subsection{Elastic mean free paths}
The elastic mean free path ($\ell_{el}$) is a key quantity in mesoscopic transport. Hereafter we focus on the case of short range disorder, that allows us to illustrate common properties of transport length scales in graphene-based low dimensional materials. The Anderson disorder model is the most generic case for investigating localization phenomena in low dimension. In this model, the onsite energies of $p_{z}$ orbitals assume random values within an interval $[-W/2,W/2]$ with a given probability distribution. Hereafter we assume a uniform probability distribution, i.e. ${\cal P}=1/W$.

In a situation of weak disorder, within the Born approximation scheme, $\ell_{el}$ can be derived at a certain degree of approximation, thus enabling the possibility to extract an analytical expression. The simplest approximation for 2D graphene can be derived as follows. The total density of states can be written as
\beq
\rho(E)=(\sqrt{3}a^{2}/2\pi) |E|/(\hbar v_{F})^{2} .
\eeq
As a result, writing $\ell_{el}=v_{F}\tau$, and using a simple Fermi Golden Rule approach for the elastic scattering time $\tau$ ($\tau^{-1}=(2\pi/\hbar)\rho(E_{F})W^{2}/12$), we obtain
\beq
  \ell_{el}\sim (\gamma_{0}/W)^{2}a|\gamma_{0}|/|E|,
\eeq
which diverges when $|E|\to 0$. This crude estimation pinpoints a difficulty in calculating transport length scales when the Fermi level lies close to the Dirac point. A numerical calculation within the Kubo approach allows the evaluation of $\ell_{el}$ at a quantitative level in 2D disordered graphene with Anderson scattering potential. In Fig. \ref{FIGmob} (inset) (adapted from \cite{LherbierNWs}), $\ell_{el}$ is shown for $W=1,1.5,2,2.5$ in $\gamma_0/2$ unit and ranges from several tens of nanometers down to a few nanometers close to the Dirac point (for $W\approx 3.4$ eV).

In quasi-1D systems (such as CNTs and GNRs), scattering angles are restricted to two cases: Forward scattering events at zero angle lead to momentum relaxation but do not affect the elastic transport length scale. Only the backscattering events at an angle of $\pi$ are taken into account for the derivation of the elastic mean free path $\ell_{el}$. 
By using the Anderson disorder model, White and Todorov derived an analytical formula for the elastic mean free path ($\ell_{el}$) close to charge neutrality point \cite{Whitetodorov}. For armchair metallic nanotubes, they obtained
\beq
 \ell_{el}=18\sqrt{3}a_{cc}(\gamma_{0}/W)^{2}N,
\eeq
showing that, for a fixed disorder strength, $\ell_{el}$ will upscale linearly with the nanotube diameter, a property unique to these systems and pinpointing long ballistic systems. Triozon and co-workers \cite{Triozon} numerically confirmed such prediction and further reported on the strong energy dependence of $\ell_{el}$ close to the onsets of new subbands. Similarly, $\ell_{el}$ was derived in metallic N-aGNRs  \cite{ARE_NL7} as
\beq
 \ell_{el}=12(\gamma_{0}/W)^{2}(N+1)a_{cc}.
\eeq
Therefore, both low dimensional carbon systems show a mean free path that diverges with increasing diameter or ribbon width for a fixed disorder strength $W$. Note however that only armchair nanotubes really display 1D massless Dirac fermions close to the charge neutrality point, since there is a gap opening in all GNRs due to edge boundary conditions.

\begin{figure}[ht]
 {\resizebox{8cm}{!}{\includegraphics{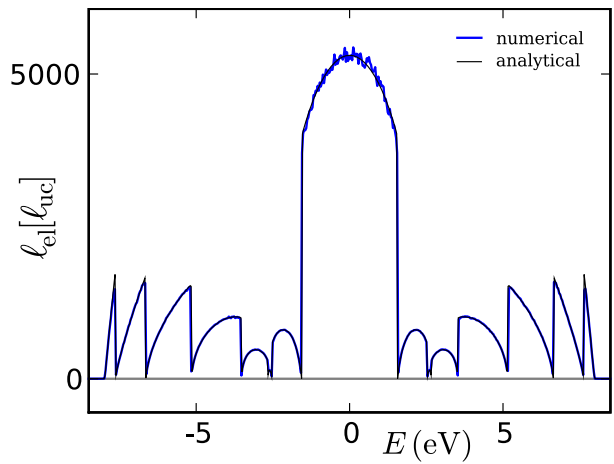}}}
 \caption{\label{l_el-perturbative}The elastic mean free path $\ell_{el}$ in a (5,5) CNT with Anderson-disorder of strength $\sigma_{\varepsilon} = 0.1 \tmop{eV}$. The numerical results are obtained from the transmission of a single disordered unit cell, averaged over 4000 samples as $\ell_{el} = L / \left( N_{\tmop{ch}} / \left\langle T \right\rangle - 1 \right)$. The analytical result is obtained via Eq.~(\ref{l_el-general}) using $g (E)$ of a disorder-free system.}
\end{figure}

An expression for $\ell_{el}$ that holds at arbitrary energies can be derived from the Fermi Golden Rule \cite{NemecGNR}
\begin{eqnarray}
  \ell_{el} & = & \ell_{uc}
  N_{\mathrm{\tmop{ch}}} \left[ \pi^2 \sum_i^{\mathrm{\tmop{uc}}}
  \sigma_{\varepsilon_i}^2 g_i^2 \left( E \right) \right]^{- 1} 
  \label{l_el-general}
\end{eqnarray}
with $\ell_{uc}$ the length of the unit cell and the sum running over all orbitals $i$ within one unit cell. In this form, the expression can be applied to arbitrary quantum wires, including GNRs, where it also covers the special case of edge disorder by making $\sigma_{\varepsilon_i}^2$ dependent on the orbital number $i$.

Neglecting multiple scattering, the elastic mean free path $\ell_{e}$ and the diffusive transmission $T_{\mathrm{\tmop{diff}}}$ are defined in terms of the {\tmabbr{LDoS}} $g_i \left( E \right)$ of the disorder-free system. Likewise, $N_{\mathrm{\tmop{ch}}}$ is defined by the leads, where it follows an exact integer step function. Near band edges, this diffusive transmission is discontinuous, as can be confirmed numerically to arbitrary precision, computing it as the sample average $\left\langle T \right\rangle$ of the transmission of many disorder configurations, as displayed in Fig.~(\ref{l_el-perturbative}).

For stronger disorder and extended disordered regions, the elastic mean free path is no longer a purely perturbative quantity due to the fact that it depends on the density of states that has to include non-perturbative effects near van Hove singularities. We can, however, retain Eqn.~(\ref{l_el-general}) by simply including the non-perturbative effects in $g_i (E)$ using Eqn.~(\ref{Eqn:g(E)}). Furthermore, the number of channels $N_{\tmop{ch}}$ must also take into account the non-perturbative effects near band-edges, which can be achieved by including the self-energy term Eqn.~(\ref{self-energy}) in the calculation of the transmission of through a cross section of an infinite quantum wire as described in Ref.~{\cite{NemecGNR}}. Incorporating these effects, we obtain the elastic mean free path in an infinitely long disordered quantum wire as displayed in Fig.~\ref{l_el-Huegle}.

\subsection{Quantum interference effects and localization phenomena in disordered graphene-based materials}
The knowledge of the mean free path $\ell_{e}$ in quasi-1D systems is crucial since it allows the identification of the frontier between the ballistic and the diffusive propagation of wavepackets in weakly disordered systems. Assuming that the transport regime remains coherent, a new class of scattering paths will yield an important contribution to the resistance, known as the weak localization correction, which eventually turns the metallic state to an insulating one \cite{AND_PR124,ABR_PRL42,LEE_PRL47,LEE_RMP57}. The localization length $\xi$ is the other physical length scale that defines such a transition, where the conductance decays exponentially with the system length as $G\sim G_{0}\exp(-L/\xi)$.

Weak localization phenomena have been observed in multiwalled carbon nanotubes with diameter ranging from $\sim 3-20$ nm \cite{ROC_PRL87,FED_PRL94,STR_SST21,FED_NL7,LAS_PRL98,STO_NJP9}. Similarly, weak localization has been recently clearly unveiled in GNRs with widths in the order of $\sim 200-500$ nm \cite{WL-ribbons}. Additionally, transition to weak anti-localization has been reported \cite{WL-ribbons,MO_PRL97,WU_PRL98} as well. Weak anti-localization (WAL) in graphene-based materials is argued to originate from some pseudospin-induced sign change of the quantum correction, similar to what is observed in systems with strong spin-orbit coupling \cite{MCC_PRL97,OST_PRB74,MOR_PRL97,KHV_PRB75}. For the same kind of disorder potential, preservation of pseudospin symmetries might jeopardize for GNRs with small width. We can expect that the effect of edge disorder and intrinsic defects (topological, vacancies, adsorbed impurities, ...) will play an increasing role as GNR width decays from $\sim 20$ nm down to $\sim 5$ nm.

It is thus genuinely important to evaluate the varying effects of disorder on quantum transport as the dimensionality or symmetries are changed. It is worth stressing that, in order to unveil weak localization effects, an external magnetic field is generally applied to tune the intensity of quantum corrections. Indeed, the phase of the quantum wavefunction is modified by the magnetic field (through a term giving the circulation of the potential vector along the scattering path), which reduces the probability of return to the origin and enhances the conductance (this phenomenon is known as the negative magnetoresistance effect). However, for low dimensional systems such as carbon nanotubes, it was demonstrated that magnetic field has also severe consequences on the band structure, so that the resulting magnetofingerprints in localization regimes become more complicated \cite{ROC_PRL87,FED_PRL94,STR_SST21,FED_NL7,LAS_PRL98,STO_NJP9} to follow. This will also apply to magnetoresistance effects in GNRs with width $\leq 10$ nm.

\begin{figure}[htp]
  \centering
  \includegraphics[width=8cm]{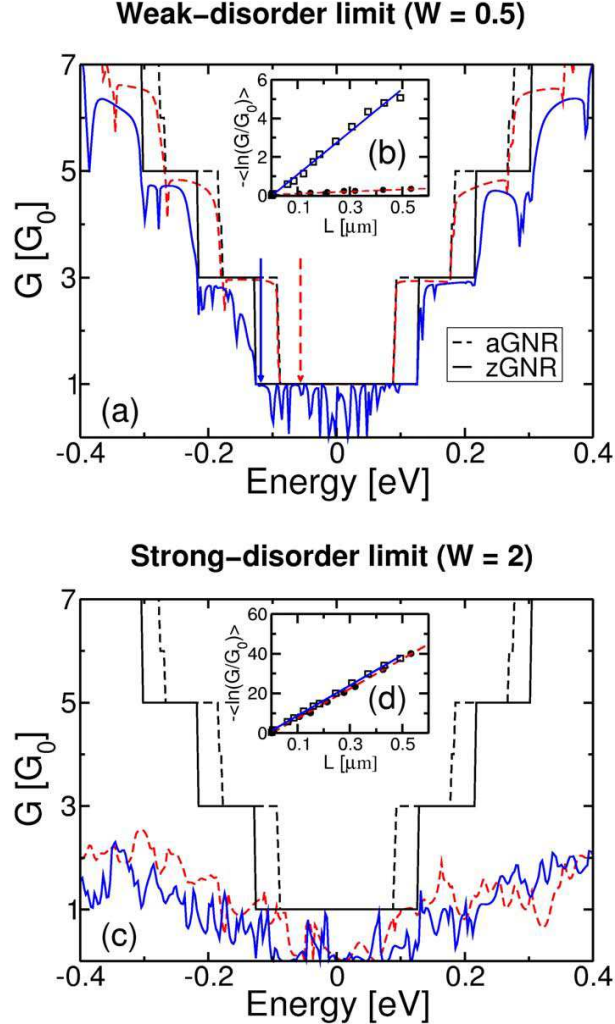}
  \caption{(a) Conductance for a single disorder configuration of a zigzag (solid blue line) and an armchair (dashed red line) GNR with width $\sim 20$ nm and for $W=0.5$. Black lines correspond to ideal zigzag (solid line) and armchair (dashed lines) ribbons. (b) Configuration averaged (over $\sim 400$ samples) normalized conductance as a function of GNR length for both zigzag and armchair GNRs. The solid blue (dashed red) arrow shows the energy at which the calculations for the zGNR (aGNR) have been performed. (c) and (d) Same informations as for (a) and (b) but for a larger disorder strength ($W=2$). Adapted from [Lherbier, A.; Biel, B.; Niquet, Y. M.; Roche, S. {\it Phys. Rev. Lett.} {\bf 2008}, 100, 036803].}
  \label{FIG4}
\end{figure}

Let us illustrate the quantum localization effects in GNRs. As seen in Section II-B, zigzag type GNRs display very peculiar electronic properties, with wavefunctions sharply localized along the ribbon edges for energies in the vicinity of the CNP.  Using a conventional Landauer-B\"uttiker approach \cite{Datta,ARE_NL7,GUN_APL90,GUN_PRB75}, one can explore the scaling properties of the quantum conductance of these systems. In Fig.~\ref{FIG4}(a) and Fig.~\ref{FIG4}(b), the energy-dependent conductance for both zGNR and aGNR of width $\sim 20$ nm are shown for pure, weak disorder ($W=0.5$) and strong disorder ($W=2$) limits.

In the weak disorder limit ($W=0.5$, Fig.~\ref{FIG4}(b)), aGNRs appear to be much less sensitive to disorder effects than zGNRs with the same width. As can be seen in Fig.~\ref{FIG4}(c) and Fig.~\ref{FIG4}(e), the averaged normalized conductances are exponentially damped, following $\langle\ln G/G_{0}\rangle\sim L/\xi$, where an average over $\sim 400$ different disorder configurations has been performed. From these calculations, $\xi$ is found to be up to two orders of magnitude smaller in zigzag than in armchair ribbons in the low disorder limit ($W=0.5$), for an energy value far from the close vicinity of the CNP (following \cite{ARE_NL7,GUN_APL90,GUN_PRB75}). In contrast, for disorder strength as large as $W=2$ (Fig.~\ref{FIG4}(d)), the localization lengths for both types of ribbons are almost equal, showing that edge symmetry does not play any role. This result can be understood by the lower transport dimensionality in the case of zigzag edge symmetry, driven by more confined wavefunctions \cite{NAK_PRB54,WAK_PRB59,WAK_PRB64,PER_PRB73,MUN_PRB74,ZHE_PRB75}.

An important result of mesoscopic physics is that there exists a fundamental relationship between $\ell_{el}$ and $\xi$ known as the Thouless relation \cite{Thouless}. In a strictly 1D system, Thouless \cite{Thouless} demonstrated that
\beq
 \xi=2\ell_{el},
\eeq
whereas for quasi-1D systems (with $N_{\perp}(E)$ conducting channels), the relation was generalized as
\beq \label{eq:xsi}
 \xi(E)=[\beta(N_{\perp}(E)-1)/2+1]\ell_{el}(E),
\eeq
with $\beta$ a factor dependent on the time-reversal symmetry \cite{RMPBeenaker}. Avriller and co-workers \cite{Avriller} recently confirmed
numerically such relation by studying chemically doped metallic carbon nanotubes.

\begin{figure}[h]
 {\resizebox{8cm}{!}{\includegraphics{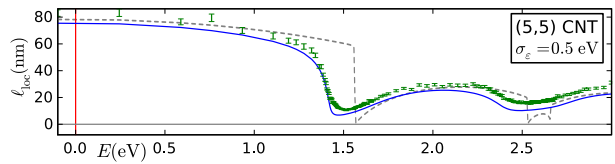}}}
 \caption{\label{fig:eloc-analytic}Localization length of a (5,5) CNT. Dashed line: value obtained from Eqn.~(\ref{eq:xsi}) by na\"{\i}vely using the perturbative values for $\ell_{el}$ and $N_{\perp}$ based on the self-energy and DoS of the disorder-free system. Solid line: same value based on the non-perturbative quantities including disorder in the self-energy and DoS. The numerical values indicated by error bars are obtained from logarithmic sample averaging. Adapted from [Nemec, N., Richter, K.; Cuniberti, G. {\it New J. Phys.} {\bf 2008}, 10, 065014].}
\end{figure}

As it was shown in Ref.~{\cite{NemecGNR}}, Eqn.~(\ref{eq:xsi}) still holds to good precision near van Hove singularities on the condition that the non-perturbative effects of multiple scattering are correctly included in the calculation of the DoS and the number of channels $N_{\perp}$ as it was described before for the elastic mean free path. As can be seen in Fig.~\ref{fig:eloc-analytic}, this correction greatly improves the agreement with the true value obtained via sample-averaging. The remaining deviation is predominantly caused by the fact that the different conduction channels have very different velocities and thereby very different elastic mean free paths, whereas the Eqn.~(\ref{eq:xsi}) is based upon the assumption of equivalent channels with one common elastic mean free path.

\begin{figure}[ht]
 {\resizebox{15cm}{!}{\includegraphics{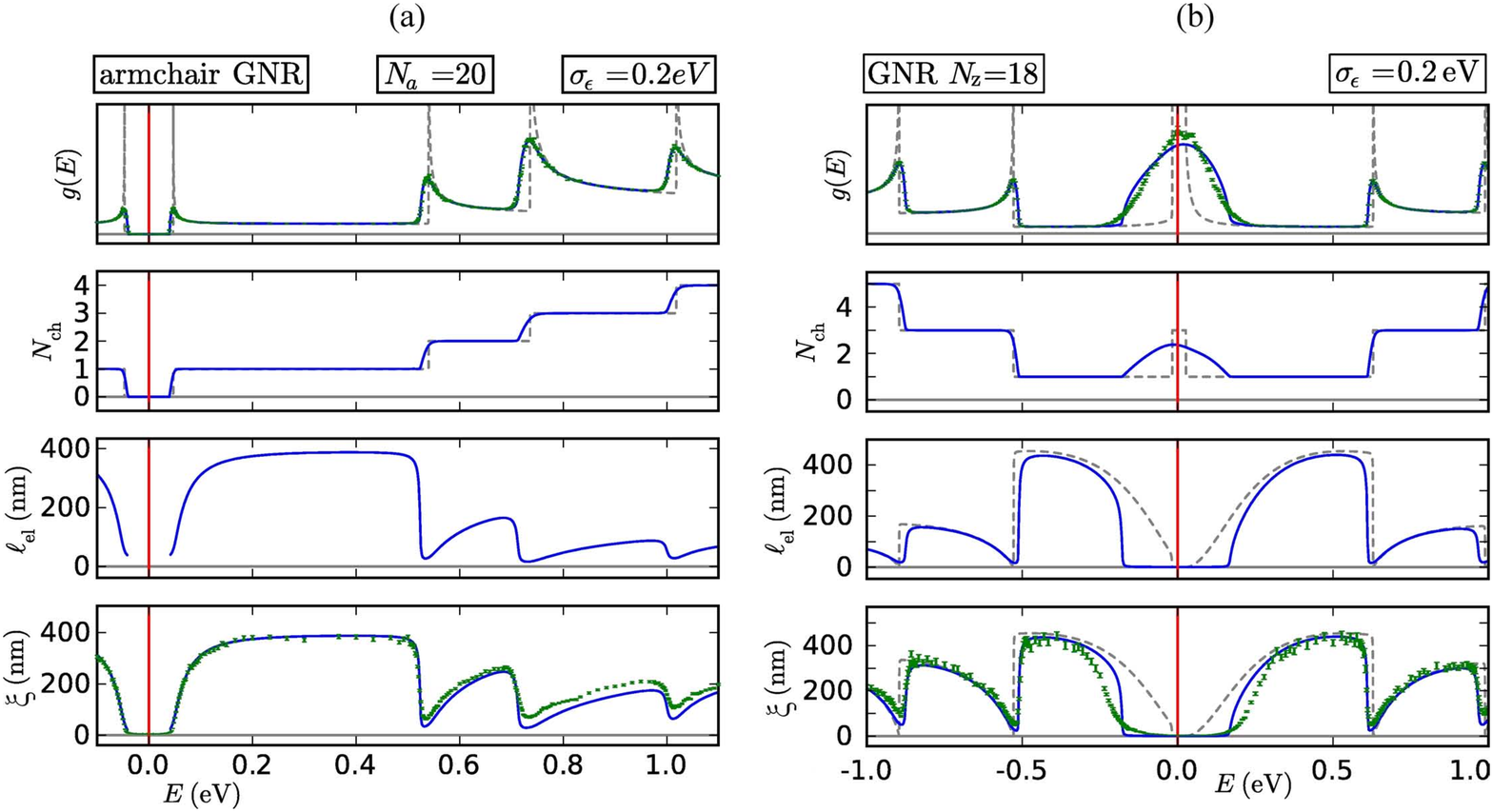}}}
 \caption{\label{NN:GNRa20} (a) Density of states $g(E)$, number of channels $N_{\tmop{ch}}$, elastic mean free path $\ell_{el}$ and localization length $\ell_{\tmop{loc}}$ of an armchair GNR of width $N_a = 20$ under the influence of Anderson disorder ($\sigma_{\varepsilon} = 0.2 \hspace{.2em} \mathrm{\tmop{eV}}$). Dashed lines: $g \left( E \right)$ and $N_{\tmop{ch}}$ of the clean system. Solid lines: $g\left( E \right)$ from Eq.~(\ref{Eqn:g(E)}), $N_{\tmop{ch}}$ obtained as transmission through cross section through infinite system with self-energy Eq.~(\ref{self-energy}) and lengths obtained from these. Data with error bars: values obtained numerically by averaging over $\sim 180$ samples of length $2000$ to $20000 \ell_{uc}$. (b) The same for a zigzag GNR of width $N_z = 18$. The most prominent feature is the disorder-broadened peak in the DoS caused by the edge state that causes a strong reduction in the elastic mean free path and the localization length. The analytical values fail to describe the true shape of this feature because the expressions neglect the difference in the various channels. Adapted from [Nemec, N., Richter, K.; Cuniberti, G. {\it New J. Phys.} {\bf 2008}, 10, 065014].}
\end{figure}
The expressions for the DoS and the localization length obtained and demonstrated before for carbon nanotubes are general enough to apply to graphene nanoribbons as well. The data in Fig.~\ref{NN:GNRa20}(a) demonstrates that the values obtained analytically do indeed match the results of numerical sample averaging. For semiconducting CNTs and GNRs, the behavior near the gap is correctly reproduced, showing a slight disorder-induced reduction of the gap-width.

As can be seen in Fig.~\ref{NN:GNRa20}(b), the expressions still hold for zigzag GNRs over most of the energy spectrum, and even qualitatively describe the effects of the edge state around the Fermi energy: the extremely high DoS of the Fermi energy is smeared out in the energy range and the localization length is drastically reduced not only in the narrow region of the low-dispersive edge state itself, but - due to the disorder-induced broadening in energy - in an extensive region around the Fermi energy. However, the expressions fail to describe the actual shape of the energy-broadened peak in the DoS, as well as the flanks of the suppressed region in the localization length. These strong deviations can be traced to the assumption of equivalent conduction channels that is made in Eqn.~(\ref{eq:xsi}). Near the edge state, this assumption fails completely.

\subsection{Edge disorder in GNRs}
In contrast to two-dimensional graphene and carbon nanotubes, graphene nanoribbons are subject to chemical passivation and roughness at the edges. In some cases, the nature of the chemical groups that passivate the edges (usually H) can be determined experimentally, thus enabling a certain control of the ribbon. Unfortunately, the spatial regularity of the edges is much more difficult to achieve and the state-of-the-art etching techniques cannot avoid roughness. Direct or indirect evidences of edge disorder have been observed in many and different graphene samples, independently of the technique exploited to fabricate them.

In the literature, the edges of graphene sheets exfoliated by cleavage of highly oriented pyrolitic graphite have been directly investigated by scanning tunneling microscopy (STM) in association with scanning tunneling spectroscopy (STS) \cite{NII_PRL73,ENO_DRM16,KOB_PRB73,ENOK_IRPC07} and by direct contact atomic force microscopy (AFM) in \cite{BAN_PRB72,BAN88}. Micro-Raman spectroscopy has also proved to be a valuable tool for studying the armchair or zigzag orientation of the ribbon edges locally \cite{CAN_PRL93}. 

 In all these experiments, the structure of the edges turned out to be very irregular, with alternation and mixing of zigzag and armchair terminations, protrusions and dents or more complex structures. In general, armchair segments are considerably longer than zigzag fragments, thus evidencing a lower stability for the latter. The measurement of the transport properties of lithographically etched ribbons \cite{HAN_PRL98,CHE_PE40} also provides indirect evidences of the edge roughness. In particular, hints of inactive edge regions and dependence of the maximum resistivity on the ribbons width have been related to a possible disorder on the edges.

Similar conclusions are drawn for ultrathin epitaxial graphene grown on silicon carbide crystals \cite{BER_SCI312,HEE_SSC143}. In this case, an indication of edge disorder (due to roughness or chemisorbed molecules) comes from the lower-than-expected number of conductive modes, due to inactive edge regions, and from the behavior of weak localization.  

Innovative techniques for fabrication and etching of graphene ribbons have proved to reduce the edge irregularities considerably, besides allowing the realization of ultranarrow (few nm) structures. The chemical technique developed by Dai and co-workers \cite{XIA_SCI319} enables the formation and selection of long and ultrasmooth nanoribbons. Despite the high spatial regularity, field-effect transistors based on these systems \cite{WAN100} have shown a non-negligible scattering related to disorder at edges. The very recent STM lithographic technique developed by Tapaszt\'o and co-workers \cite{TAP}, allows the patterning of ultranarrow structures with the possibility of choosing orientation and width with almost atomic precision. Nevertheless, STM measurements on few nm wide ribbons have revealed irregular oscillations in the electronic density of states, thus suggesting the possible presence of edge disorder and its importance for very narrow ribbons.   

\begin{figure}
\includegraphics{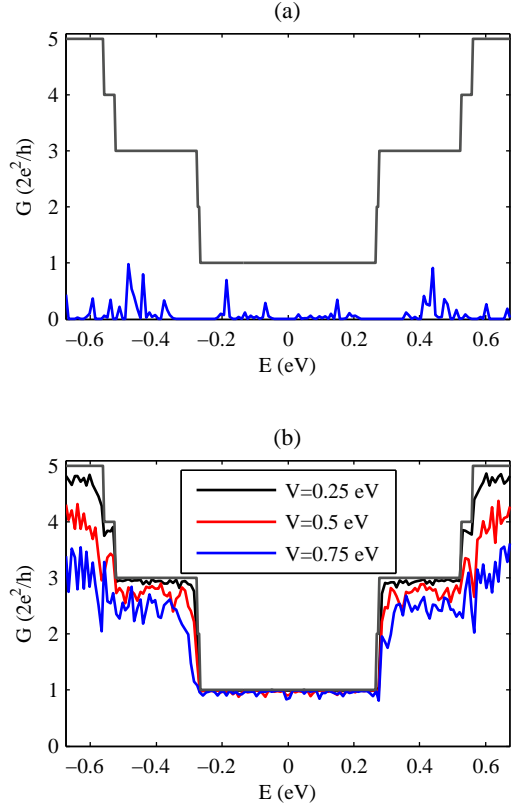}
\caption{(a) Differential conductance of a metallic armchair ribbon (N=53) with a $\sim$10\% of vacancies on the edges. Couple of atoms have been removed on the most external rows over a length of about 1 $\mu$m. For the calculation, we adopted a simple tight-binding Hamiltonian with zero site energy and hopping energy $\gamma_0=-2.7$ eV. (b) Differential conductance for the same system with Anderson disorder on the edges. The energy of the edge sites varies randomly in the range $\pm V$ with $V=$0.25, 0.5 and 0.75 eV over a length of about 1 $\mu$m. }
\label{fig_armchair}
\end{figure}

The literature proposes several and different edge disorder models for graphene nanoribbons. The basic idea to start from a tight-binding Hamiltonian for a clean and regular ribbon (in general with hydrogen passivation on the edges), and then adding \cite{GUN_APL90} or removing \cite{ARE_NL7,QUE92,EVA,MUC} carbon atoms at the edges, or varying the width of the system \cite{MAR} to account for roughness, or introducing Anderson disorder \cite{LI77}. The most proper way of adding/removing atoms at the edges avoids final configurations that might cause steric problems. In practice, only H-C-C-H groups can be removed or added at the edges of an armchair ribbon, while particular care must be adopted when the disorder goes deeper than the first row in zigzag ribbons. [Not all the authors take care of these constrains.]     

\begin{figure}
\includegraphics{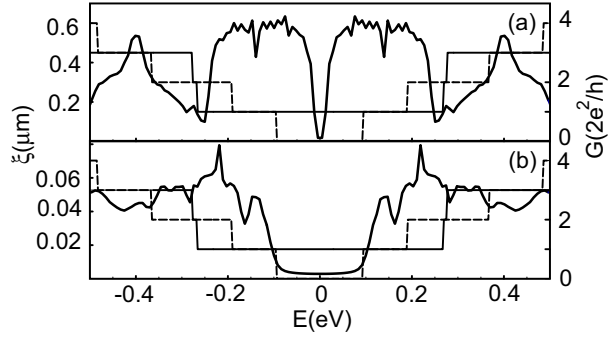}
\caption{One half the amplitude of the localization length for a metallic aGNR with $N=53$ for a 5\% (a) and a 50\% (b) concentrations of defects at the edges. This figure is taken from [Areshkin, D. A.; Gunlycke, D.; White, C. T. {\it Nano Lett.} {\bf 2007}, 7, 204], by courtesy of C.T. White.}
\label{ARE7_fig6}
\end{figure}

The transport properties of metallic aGNRs have been investigated in the presence of both roughness \cite{ARE_NL7,MUC} and Anderson disorder \cite{LI77} on the edges. In the case of vacancies on the edges, the differential conductance of the system is considerably reduced and even a weak disorder on the two external rows induces a localization process, see Fig. \ref{fig_armchair}(a). The localization length turns out to be particularly low around the neutrality point even for a 5\% of vacancies. This is attributed to a small gap opening in correspondence of the crossing of the two bands around E=0 \cite{ARE_NL7} and whose width is inversely proportional to the width of the ribbon \cite{MUC}. When increasing the level of disorder on the two most external rows, the system can be envisioned as a sequence of metallic and semiconducting fragments and the localization length decreases considerably within the energy range that corresponds to the semiconductor gap, see Fig. \ref{ARE7_fig6}. Even outside this region, the conduction ability is seriously jeopardized, with a localization length of few tens of nm. [Similar results have been also recently obtained by Evaldsson et al. \cite{EVA}.] 

The effect of sequences of large conducting and semiconductor armchair fragments has been investigated in detail by Martin et al. \cite{MAR}. The length of each metallic fragment is such to preserve the band structure of the corresponding infinitely extended ribbon. From this perspective, it is possible to obtain an effective tight-binding Hamiltonian, where the eigenfunctions of the metallic fragment play the role of ``orbitals'', with equispaced site energies, and the hopping energies between two subsequent fragments are related to the properties of the semiconductor fragment in between. The result is a one-dimensional impurity band insulator with conductivity
\beq \sigma \approx e^{-2\sqrt{\alpha E_g/T}} \ \ \ \ {\rm for} \ T<T^*\eeq
where $L_{\rm av}>W$ is the average length of the metallic grains, $E_g$ is the energy gap of the semiconductor fragments, $T$ is the temperature, $T^*\approx |\gamma_0| D/L_{\rm av}^2$ (with $D$ the width of the ribbon) and $\alpha$ is a numerical coefficient of the order of 1.

\begin{figure}
\includegraphics{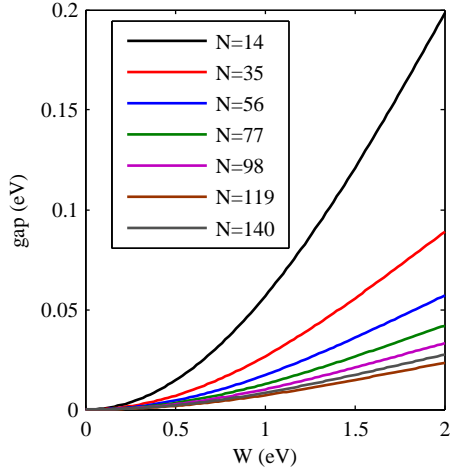}
\caption{Gap between the highest valence band and the lowest conduction band in the presence of a constant potential $W$ on the two edges of the armchair ribbon for $N=$14,35,56,77,98,119,140. The larger is the strip, the smaller is the gap.}
\label{fig_gap}
\end{figure}

In the case of weak Anderson disorder on the edges, the conductance of metallic aGNRs is only slightly affected, in particular within the energy region that corresponds to a single channel, see Fig. \ref{fig_armchair}(b). This result can explained by considering the high kinetic energy of the states around the charge neutrality point \cite{LI77}, which are therefore scarcely affected by potential fluctuations. Just to give an idea of the low sensitivity of these states to disorder, let us consider a constant potential on both edges. It turns out that the structure of the highest valence band and the lower conduction bands does not change around the charge neutrality point but for a small shift in energy and the rising of an energy gap whose width is always small and considerably suppressed for larger ribbons, see Fig. \ref{fig_gap}. The linear energy dispersion is thus preserved and the gap is almost negligible because the states that correspond to the bands around the neutrality point are spread all over the section of the system and thus the effect of the edge potential is weaken by the averaging over the chains.

The two types of disorder, always confined to the two outer rows of the metallic aGNRs, have different consequences. However, there is no inconsistency in this, since the nature of the perturbation is completely different in the two cases. Roughness tends to introduce semiconductor islands and then a gap. Weak Anderson disorder does not perturb the structure of the energy bands close to the neutrality point, therefore the backscattering is limited, at least for wide ribbons.

In the case of roughness for semiconductor aGNRs, the system goes toward a strong localization regime, as in the case of metallic ribbons. However, the disorder induces states in the region of the energy gap \cite{GUN_APL90,QUE92}, thus allowing current tunneling for not very long systems. Therefore, a compromise between the width and the length of the ribbon must be found in order to keep the semiconductor behavior of the strip. In particular, the ribbon must be short enough to preserve the extension of the states just outside the gap all over its length and long enough to prevent tunneling through the states induced within the energy gap. As reported at the end of the section, this can considerably affect the $I_{\rm on}/I_{\rm off}$ ratio for GNRs-based field effect transistor \cite{YOO91,FIO,BAS92}. 
  
In the case of zigzag ribbons, the almost dispersionless edge states around the neutrality point play a key role in determining the properties of the disordered system. In the literature, different behaviors have been observed depending on the specific model of edge disorder, with results not always consistent. Areshkin and co-workers \cite{ARE_NL7} have found that zGNRs are much less sensitive to edge disorder than aGNRs, especially in the energy region where only one conductive mode is active. For reasonably wide strip and for an erosion of 50\% C atoms on the eight outer rows, the localization length turns out to be of the order of 10 $\mu m$ (at least ten times the value for an aGNR with a much weaker disorder). To explain this, they consider that the states within the first conductive channel can be divided into two groups: almost dispersionless edge states within a very small energy range around the neutrality point $E=0$ and bulk states with energy outside this range. The width of the range is determined by the transverse width of the ribbons. The effect of roughness is mostly on the edge states, and this explains why the bulk states are not much affected by disorder. On the other hand, the conductance is expected to be much depressed around the neutrality point, where the effect of disorder is larger. This is only partially observed in the simulations \cite{ARE_NL7}, where the conductance seems anyway to keep a strong resistance to roughness with respect to the armchair case. This behavior is not understood and would deserve further investigation.

A rather different conclusion is obtained by Querlioz and co-workers \cite{QUE92}, who evaluated and analyzed the wavefunction and the density-of-states of zigzag ribbons. From these data the mobility edge is extracted. Their conclusion is that no ribbon has zero mobility edge and that the resistance to edge disorder for zGNRs vanishes as soon as more than one edge row is eroded. In this case, the roughness induces an Anderson insulator behavior independently of the zigzag or armchair orientation of the ribbon. This result, in agreement with other recent calculations \cite{MUC}, seems to be in striking contrast with \cite{ARE_NL7}. More detailed analysis focusing on specific disorder models could unravel this ambiguity.

\begin{figure}
	\includegraphics{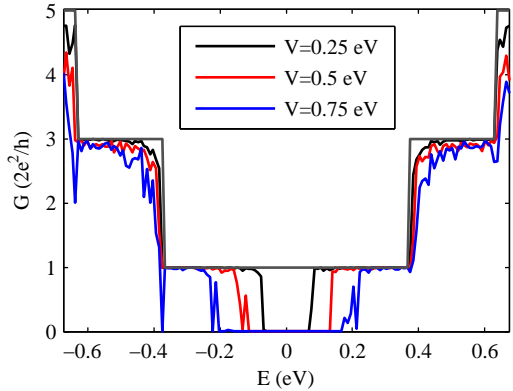}
	\caption{Differential conductance of a zigzag ribbon (N=32) with Anderson disorder on the edges. The energy of the edge sites varies randomly in the range $\pm V$ with $V=$0.25, 0.5 and 0.75 eV over a length of about 1 $\mu$m.}
	\label{fig_zigzag}
\end{figure}

Weak Anderson disorder on the edges leads to the opening of a gap in the conductance of zigzag ribbons \cite{LI77}. Some results are reported in Fig. \ref{fig_zigzag}. A weak potential on the edges has a deep impact on the electronic spectrum and then on the transport properties around the CNP. The states corresponding to the two almost flat bands at $E=0$ at the borders of the first Brillouin zone ($k=\pm\pi/a$) are completely localized on the edges of the ribbon. Therefore, a potential on the edges moves (upward or downward) the energy of these states. As a consequence, the electrons are subject to backscattering, especially in the region where most of the onsite disorder energies are concentrated, i.e. around the neutrality point. This induces a gap in the conductance, whose width increases with the strength of the potential.

We conclude this section by giving a look to the possible effects of edge roughness on the behavior of graphene-based (field-effect or Schottky-barrier field effect) transistors \cite{YOO91,FIO,YON_IEEE55,BAS92,OUY_APL92}. These systems are usually investigated with the help of Schr\"odinger-Poisson solvers, which allow the self-consistent evaluation of the electrostatic potential. The results presented in the literature agree that edge disorder reduces the $I_{\rm on}$ current and increases the $I_{\rm off}$ current, thus making their ratio worse. The decreasing of $I_{\rm on}$ is due to quantum transport effects, as seen before, and to self-consistent electrostatic effects that compensate the vacancies with an accumulation of charge on the atoms close to it. The increasing of the leakage current is due to the tunneling of electrons through the states induced in the band gap region or through conducting zigzag fragments. The degrading of the $I_{\rm on}/I_{\rm off}$ ratio is thus a serious issue for the efficiency of transistors based on very small ribbons. On the other hand, larger ribbons would reduce the gap tremendously. 
A recent theoretical paper by Ouyang et al. \cite{OUY_APL92}, in agreement with experiments \cite{WAN100}, shows that the presence of optical phonons (OPs) reduces the effect of edge irregularities considerably. In fact, electrons loose much energy by emitting an OP and they can hardly come back to the source after being reflected by the edges. As a consequence, elastic backscattering is only relevant when occurring close to the source, where electrons keep their energy before phonon emission.

\subsection{Minimum conductivity and charge mobility in graphene based systems}
\label{section_mobilities}
In semiconducting materials (such as silicon nanowires \cite{LherbierNWs}), charge mobility is a very important quantity to assess the transport efficiency of the system and the corresponding device performances. By definition, the charge mobility reads:
\begin{equation}
      \mu(E) =\sigma_{\rm sc}(E)/en(E), 
\label{eqmu}
\end{equation}
where
\begin{equation}
    \sigma_{\rm sc}=e^{2}\rho(E)v(E)\ell_{el}
\end{equation}
is the semiclassical conductivity deduced from the Einstein formula, with $\rho(E)$ the DoS, $n(E)$ the charge density at energy $E$, $\ell_{el}$ the elastic mean free path, and $e$ the elementary charge. Close to the charge neutrality (or Dirac) point, the measured experimental conductivity of various samples was found to range within $\sim 2-5 e^{2}/h$ although the charge mobility was changing by almost one order of magnitude \cite{graphene,BER_SCI312,ZHA_N438,ZHA_PRL96,GEI_NM6,OEZ_APL91,OEZ_PRL99,JIA_PRL99,NOV_SCI315,WU_PRL98}. This effect has been attributed to the change of charge density due to the doping from the substrate and/or contacts.

On the theoretical side, the calculation of the Kubo conductivity for 2D graphene with short range disorder, and within the self-consistent Born approximation, yields $\sigma^{\rm min}_{xx}=4e^{2}/(\pi h)$ ($h$ the Planck constant) for the two Dirac nodes \cite{SCBA}, which is typically $1/\pi$ smaller than all the experimental data. Numerical calculations using the Kubo formula confirm such prediction \cite{NOM_PRL96,PRLlherbier1}. Amazingly, as discussed in section \ref{pseudo}, this value also comes out in a completely ballistic transport regime, as a contact effect.

Differently, by assuming that the Dirac fermion scattering is dominated by Coulomb scattering from ionized impurities near the graphene plane, Nomura and MacDonald \cite{NOM_PRL96} could numerically reproduce the low energy dependence of the measured charge conductivity, and found that $\sigma^{\rm min}_{xx}\sim e^{2}/h$ close to the Dirac point, in better agreement with most experiments. In \cite{NOM_PRL96}, the authors used the full quantum approach of the Kubo formula, describing the long-range disorder effects via a screened Coulomb potential, and performing a finite-size scaling analysis. 

Other calculations within the semi-classical Boltzmann approach or the Landauer approach have also reported on the effect of screened Coulomb potential on charge conductivity \cite{LEW_PRB77,HWA_PRL98} with similar conclusions, and although the existence of a true universal minimum conductivity could not be rigorously answered, an interpretation in terms of saturation of the conductivity due to charged impurity induced inhomogeneities, occurring at low densities, was proposed \cite{HWA_PRL98}. 

These results suggest that the intrinsic disorder in graphene could be of electrostatic (Coulomb scattering) nature, likely due to charges trapped in the oxide. One also notes that the effect of graphene plane deformation modes (known as ripples), frozen when the exfoliated layer is deposited onto an oxide layer, has been investigated by introducing an effective random gauge potential \cite{MOR_PRL97,NOM_PRL100,NETO}. The authors found that a temperature independent minimal conductivity will take place, with full suppression of localization effects owing to the absence of intervalley scattering processes, in agreement with temperature-dependent experiments \cite{MOR_PRL100}. 
 
Concerning the contribution of quantum interference effects and the transition to an Anderson type localized regime, the issue is still controversial. In presence of strong intervalley scattering processes, which is best realized for short range disorder potential, conventional weak localization phenomena have been predicted \cite{SUZ_PRL89,MCC_PRL97} and observed \cite{WL-ribbons}. The preservation of the pseudospin symmetry also allows for the manifestation of a spectacular symmetry-dependent anti-localization effect \cite{SUZ_PRL89,MCC_PRL97,WL-ribbons}. The transition to a localized Anderson type regime is more debated \cite{SUZ_JPSJ75,ALE_PRL97}, with to date no experimental evidences in such material. The question is how quantum interferences effects and localization phenomena for Dirac fermions depend on the underlying disorder potential characteristics, and whether the conventional 2D scaling theory of localization \cite{LEE_PRL47,LEE_RMP57} is applicable or not in today's graphene materials.

Several authors have challenged the single parameter scaling theory of Anderson localization by studying the so-called beta function $\beta(g)=d \ln g/d \ln L$, with $g$ the dimensionless conductance and $L$ the system size. By computing the scaling behavior of $\beta(g)$, the localization versus delocalization nature of electronic states can be analyzed. In conventional two-dimensional systems, the theory predicts that all states are localized, independently of the disorder characteristics and provided time reversal symmetry is preserved \cite{LEE_PRL47,LEE_RMP57}. 

\begin{figure}[htp]
     \centering
     \includegraphics[width=8cm]{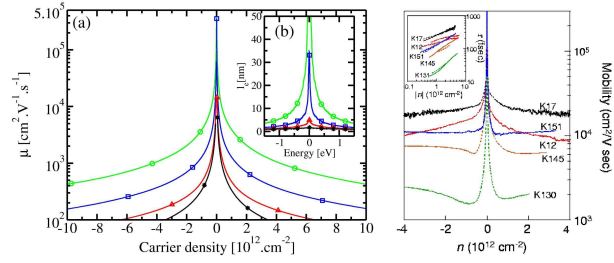}
     \caption{Left: charge mobility as a function of carrier density (a) and mean free path (b) for $W=1, 1.5,2,2.5$ (from top to bottom), from [Lherbier, A.; Persson, M. P.; Niquet, Y. M.; Triozon, F.; Roche, S. {\it Phys. Rev. B.} {\bf 2008}, 77, 085301]. Right: experimental mobility from [Tan, Y. W.; Zhang, Y.; Bolotin, K.; Zhao, Y.; Adam, S.; Hwang, E. H.; Das Sarma, S.; Stormer, H. L.; Kim, P. {\it Phys. Rev. Lett.} {\bf 2007}, 99, 246803], by courtesy of Philip Kim.}
     \label{FIGmob}
\end{figure}

Recent numerical studies claim that in presence of short range disorder (Anderson-type potential) and intervalley scattering, all states are indeed localized even for Dirac fermions \cite{PRLlherbier1}. However, it is interesting to note the typical values obtained for transport length scales. By varying the disorder strength from $W\simeq 3.4$eV to $W\sim 2$eV, the elastic mean free path was found to range from a few nm to several tens of nm close to the Dirac point (Fig.\ref{FIGmob}-left panel (b)), whereas the localization length $\xi$ given by $\xi=\ell_{e}\exp(\pi\sigma_{sc}/G_{0})$ upscaled from 20nm to 10$\mu$m (not shown here, see Ref. \cite{PRLlherbier1}). The energy-dependence of $\xi$ was also shown to be strongly driven by that of the semi classical conductivity with a minimum localization length at the Dirac point \cite{PRLlherbier1}.

In case of long range disorder, the situation is more complex. Indeed, in absence of intervalley scattering, Dirac fermions cannot be trapped by a potential well, irrespective of the well depth. This suggests the robustness of states against an insulating tendency. A different kind of scaling behavior of the conductance at the Dirac point was first proposed in \cite{OST_PRL98}. The key result was the occurrence of a quantum critical point at half filling giving some universal value of the conductivity of the order of $e^{2}/h$, so in contradiction with a localized nature giving a zero conductivity in the thermodynamic limit. Other numerical studies \cite{BAR_PRL99, NOM_PRL99} found a different scaling flow for the beta function of the Dirac model, thus indicating that all states remain delocalized whatever the strength of the underlying disorder is, but with a conductivity upscaling with length up to infinity. The situation changes in the presence of ripples, which can be described by random vector potentials. Nomura and co-workers \cite{NOM_PRL100} further deepen the scaling behavior of both transport coefficients $\sigma_{xx}$ and $\sigma_{xy}$ and found that massless Dirac fermions will exhibit a critical behavior similar to that of the quantum Hall transition point, but in absence of uniform magnetic field.

An important observation is that to date, the theoretical description of disorder in graphene layers (either deposited on a substrate or suspended in between contacts) is mostly achieved at a phenomenological level. Usually, short range and long range scattering potential are described by some onsite potential fluctuations (Anderson-type) and by Gaussian correlated potentials, respectively. Although the study of deformation modes known as ripples is at the origin of peculiar transport predictions \cite{NOM_PRL99,Guinea-PRB08}, a realistic description of Coulomb scatters is however needed to allow a true experimental exploration of localization phenomena in massless Dirac fermions. Novikov \cite{Novikov-APL07} first discussed about the possible asymmetry in the transport cross section for a Dirac electron scattering off a positively or negatively charged Coulomb impurity. Self consistent RPA-Boltzmann theory also found Coulomb scattering induced conductivity asymmetry \cite{Adam-PNAS07}, without the possibility however to tackle with localization effects. A recent theoretical study has investigated quantum coherent transport and transport scaling lengths for intentionally chemically doped (and disordered) 2D graphene layers with a realistic and self consistent description of impurity scattering potentials \cite{LHE_PRL101}. By incorporating substitutional boron (or nitrogen) impurities, elastic mean free paths, as well as the semi-classical conductivity and charge mobilities were numerically estimated by the Kubo approach. Some onset of quantum interference effects was also observed, even at the Dirac point, but this contribution was found too small to explore the possible underlying scaling behavior of the beta function, even in situation of strong doping (such as 4\%).

Finally, although conductivity is a well defined quantity, 2D graphene manifests specific properties that make the use of formula Eq. (\ref{eqmu}) somehow ill-defined. Indeed, when the energy of charge carriers approaches the Dirac point, the semiclassical conductivity remains finite, whereas $\mu\to\infty$, as the charge density $n\to 0$.

In Fig. \ref{FIGmob}, we show the result of a numerical calculation using the Kubo approach (see \cite{PRLlherbier1} for details). The evolution of $\mu$ is shown as a function of the Anderson disorder strength $W$. The energy-dependence of $\mu(E)$ and $\ell_{el}(E)$ are found to be similar. In the close vicinity of the Dirac point, the downscaling of $\mu$ with $W$ follows the Fermi Golden Rule prediction, while it diverges when approaching the Dirac point. Experimental data from \cite{KimGmob} are also shown in Fig. \ref{FIGmob} (right) for comparison. Different samples with varying quality show a similar trend, although patterns from different samples can substantially differ in shape indicating fluctuations in the disorder characteristics. 

It is clear that by further reducing the dimensionality of the graphene material, charge mobilities will be reduced. Recent transport measurements on nanoribbon-based field effect transistor show mobilities in order of $\mu \sim 100-300{\hbox{cm}}^{2}\hbox{V}^{-1}\hbox{s}^{-1}$ \cite{WAN100}, which are thus reduced in comparison with the 2D graphene measurements (reported values can be as large as a few $100.000{\hbox{cm}}^{2}\hbox{V}^{-1}\hbox{s}^{-1}$\cite{ZHA_N438,ZHA_PRL96,GEI_NM6,OEZ_APL91,OEZ_PRL99,JIA_PRL99,NOV_SCI315,WU_PRL98}). However, the lateral size reduction allows for a larger energy gap, which ensures more efficient field effect efficiency.

\section{\label{conclusions}Conclusion}
To conclude, in this review we have reported on the basics of electronic and transport properties in low dimensional carbon-based materials including 2D graphene, graphene nanoribbons and carbon nanotubes. It has been shown that although nanotubes and nanoribbons share similar electronic confinement properties due to their nanoscale lateral sizes, the effects of boundary conditions in the perpendicular direction with respect to the system axis trigger very different transport features when disorder is included. Close to the charge neutrality point, the robustness of armchair metallic nanotubes against disorder is absent in nanoribbons, which cannot be classified in the family of 1D massless fermions, owing to edge-induced gap openings. Nanoribbons with zig-zag symmetries are even more spectacularly sensitive to disorder owing to the edge-states driven lower transport dimensionality. In contrast, for charge carrier energies lying in the higher energy subbands, the properties of nanotubes and ribbons present similar features, with strong energy dependence of elastic mean free paths and localization phenomena. 

Additionally, the transition from a quasi-1D to a true 2D system results in strong damping of disorder effects, with enhanced elastic mean free paths together with strong damping of quantum interferences. In particular, the study with Anderson disorder demonstrates that even in the strongest case of short range scattering potential (with possible short range potential fluctuations as large as 1eV), the computed 2D localization lengths remain in the range of several hundreds nanometers to microns. One can thus conclude that to observe weak and strong localization regimes, the presence of edges as well as a reduced lateral size are essential factors.

Finally, the possibility to produce and control defect densities either through intentional doping or by irradiation techniques (that produce vacancies-type defects) could open spectacular avenues to explore quantum transport phenomena (including quantum Hall effects \cite{KOSH_PRB73}) in low dimensional materials, for which a realistic description of both underlying electronic structures as well as superimposed disorder potentials would be theoretically possible, allowing unprecedented exploration of experimental data at a quantitative level.

\begin{acknowledgements}
We acknowledge fruitful and enlightening discussions with Tsuneya Ando, HongJie Dai, Toshiaki Enoki, Philip Kim, Aur\'elien Lherbier, Kentaro Nomura, Rudolf A. Roemer, Riichiro Saito, Miriam del Valle and Carter T. White.
This work was partially supported by the ANR/PNANO project ACCENT, by the FP7/ICT/FET GRAND project, by the "Graphene project" of CARNOT Institute-Leti, by the European Union project "Carbon Nanotube Devices at the Quantum Limit" (CARDEQ) under contract No. IST-021285, by the Volkswagen Stiftung under Grant No. I/78 340, by the DFG Priority Program "Quantum Transport at the Molecular Scale" SPP1243 and by DAAD. Computing time provided by the ZIH at the Dresden University of Technology is also acknowledged.
\end{acknowledgements}

\end{document}